\documentclass[aps,pra,amsmath,amssymb]{revtex4}
\usepackage{graphicx,epsfig,epsf}
\usepackage{dcolumn}
\usepackage{bm,psfrag,color}

\begin{document}

\newcommand{\ri}{{\rm i}}
\newcommand{\re}{{\rm e}}
\newcommand{\ba}{{\bf a}}
\newcommand{\bb}{{\bf b}}
\newcommand{\bx}{{\bf x}}
\newcommand{\bd}{{\bf d}}
\newcommand{\br}{{\bf r}}
\newcommand{\bk}{{\bf k}}
\newcommand{\bE}{{\bf E}}
\newcommand{\bR}{{\bf R}}
\newcommand{\bM}{{\bf M}}
\newcommand{\bn}{{\bf n}}
\newcommand{\bs}{{\bf s}}
\newcommand{\tr}{{\rm tr}}
\newcommand{\tbs}{\tilde{\bf s}}
\newcommand{\rSi}{{\rm Si}}
\newcommand{\beps}{\mbox{\boldmath{$\epsilon$}}}
\newcommand{\bthe}{\mbox{\boldmath{$\theta$}}}
\newcommand{\rg}{{\rm g}}
\newcommand{\xmax}{x_{\rm max}}
\newcommand{\ra}{{\rm a}}
\newcommand{\rx}{{\rm x}}
\newcommand{\rs}{{\rm s}}
\newcommand{\rP}{{\rm P}}
\newcommand{\up}{\uparrow}
\newcommand{\down}{\downarrow}
\newcommand{\hc}{H_{\rm cond}}
\newcommand{\kb}{k_{\rm B}}
\newcommand{\cI}{{\cal I}}
\newcommand{\tit}{\tilde{t}}
\newcommand{\cE}{{\cal E}}
\newcommand{\cC}{{\cal C}}
\newcommand{\cS}{{\cal S}}
\newcommand{\clR}{{\cal R}}
\newcommand{\Ubs}{U_{\rm BS}}
\newcommand{\qq}{{\bf ???}}
\newcommand*{\etal}{\textit{et al.}}

\def\vec#1{\mathbf{#1}}
\def\ket#1{|#1\rangle}
\def\bra#1{\langle#1|}
\def\ketbra#1{|#1\rangle\langle#1|}
\def\braket#1{\langle#1|#1\rangle}
\def\idmat{\mathbf{1}}
\def\caln{\mathcal{N}}
\def\calc{\mathcal{C}}
\def\rhon{\rho_{\mathcal{N}}}
\def\rhoc{\rho_{\mathcal{C}}}
\def\tr{\mathrm{tr}}
\def\bfu{\mathbf{u}}
\def\bfmu{\mbox{\boldmath$\mu$}}

\sloppy

\title{Heisenberg-limited sensitivity with decoherence-enhanced measurements}
\author{$^{1,2}$Daniel Braun and $^{1,2,3}$John Martin}

\affiliation{$^{1}$Universit\'e de Toulouse, UPS, Laboratoire
de Physique Th\'eorique (IRSAMC), F-31062 Toulouse, France}
\affiliation{$^{2}$CNRS, LPT (IRSAMC), F-31062 Toulouse, France}
\affiliation{$^{3}$Institut de Physique Nucl\'eaire, Atomique et de
Spectroscopie, Universit\'e de Li\`ege, 4000 Li\`ege, Belgium}

\maketitle



{\bf Quantum-enhanced measurements use quantum mechanical effects in
  order to enhance the sensitivity of the measurement of 
  classical quantities, such as the length of an optical cavity. The
  major goal is to beat the standard quantum 
  limit (SQL), i.e.~an uncertainty of order $1/\sqrt{N}$, where $N$ is
  the number of quantum resources (e.g.~the number of photons or atoms
  used), and to achieve a scaling $1/N$, known as the Heisenberg
  limit. So  far very few experiments 
  have demonstrated an improvement over the SQL. The required quantum
  states are generally highly entangled, difficult to produce, and very
  prone to decoherence. Here, we show that Heisenberg-limited measurements
  can be achieved without the use of entangled 
  states by coupling the quantum resources to a common 
  environment that can be measured at least in part. The 
  method is robust under decoherence, and in fact the 
  parameter dependence of collective decoherence itself can
  be used to reach a $1/N$ scaling.}\\

Quantum mechanical noise imposes fundamental limitations on any
measurement. The best--known example is the Heisenberg uncertainty
relation, which provides a lower bound on the product of
fluctuations of two non-commuting quantized variables.  But even a
classical system parameter $x$ can in general not be measured with
arbitrary precision with a finite number of measurements due to the
statistical nature of any quantum state. A lower bound on the
uncertainty is given by the smallest $\delta x$ such that two
quantum states $\rho(x)$ and $\rho(x+\delta x)$ lead to
statistically significant differences for an optimally chosen
observable. A similar problem exists in classical statistical
analysis, where one wants to distinguish between two probability
distributions $P(x)$ and $P(x+\delta x)$, and the celebrated Cram\'er-Rao
bound sets an ultimate lower bound on the uncertainty of a measurement of $x$
based on the distinguishability of $P(x)$ and $P(x+\delta x)$
\cite{Cramer46}. That analysis has been generalized to the
quantum world \cite{Braunstein94} and has become known as
``quantum parameter estimation theory''. Recently, the theory was used to
prove lower bounds on the smallest measurable $\delta x$ for unitary time
evolution.  It was shown that if $N$ replicas of the 
quantum system evolve independently and linearly, 
for an initially separable state no uncertainty smaller than
$1/\sqrt{N}$ can be achieved, i.e.~the SQL, no matter how
sophisticated the measurement. If initially entangled states are
allowed, a $1/N$ scaling of the uncertainty is the ultimate lower
bound under otherwise identical conditions
\cite{Giovannetti04,Giovannetti06,Budker07,Goda08}.  
The use of non-classical states of light for Heisenberg limited
interferometry, notably the use of squeezed states, was proposed
theoretically already in 1981 \cite{Caves81}. So-called
NOON states have been investigated for super-resolution
\cite{Sanders89,Boto00,Mitchell04}. However, decoherence of these highly
non-classical states has so far prevented reaching an uncertainty that
scales as $1/N$ for systems with $N\gg 1$ \cite{Leibfried05,Nagata07}. In
\cite{Higgins07} entanglement-free Heisenberg-limited sensitivity of a
phase-shift measurement 
was reported for several hundred quantum resources by passing light
many times through the phase-shifter.

Decoherence arises when a quantum system interacts
with an environment with many uncontrolled  degrees of freedom, such as the
modes of the electromagnetic field, phonons in a solid, or simply a measurement
  instrument \cite{Zurek91}. Decoherence destroys quantum mechanical coherence,
  and plays an important role in the transition from quantum to classical
  mechanics \cite{Giulini96}. It becomes extremely fast for  a mesoscopic or
  even macroscopic ``distance'' 
  between the components of a ``Schr\"odinger cat''-type superposition of
  quantum states. 
Universal power laws rule the scaling of the decoherence rates in this regime
  \cite{Braun01,Strunz02}. Only recently could the collapse be time-resolved
  in experiments with relatively small ``Schr\"odinger cat''--states
  \cite{Brune96,Guerlin07}.  However, decoherence can depend very
  sensitively on the initial state and the coupling to the environment. 
  Entire decoherence-free subspaces (DFS) can exist if the coupling
  operators to the environment have degenerate eigenvalues
  \cite{Zanardi97,Braun98,Lidar98,Duan98,Braun01B}.  

We show below that a collective coupling that depends on a parameter
$x$  of $N$ quantum systems ${\cal S}_i$ to a common ``environment'' ${\cal
  R}$ can be used to measure $x$ with an uncertainty that scales as
$1/N$ with an 
  initial product state of all subsystems. The method works whether ${\cal
    R}$ is entirely under our  
  control, or a reservoir with many degrees of freedom to which we have only
  partly access, i.e.~a collective decoherence process of the ${\cal 
  S}_i$, as long as we can measure an observable of the environment. 

\section*{\large Results}\label{sec.gencase}
\subsection*{Model}
Consider $N$ quantum systems ${\cal S}_i$ coupled to a common environment 
${\cal R}$.
The hamiltonian of the total system has the form
\begin{equation} \label{Hx}
H(x)=\sum_{i=1}^N H_i+\sum_{i,\nu}S_{i,\nu}(x)\otimes R_\nu+H_R\,,
\end{equation}
where $H_i$ is  the hamiltonian of system ${\cal S}_i$, and for simplicity we
take the ${\cal S}_i$ as non-interacting. $H_R$ denotes the
hamiltonian of ${\cal R}$, which may be itself a composite quantum
system. Hamiltonian (\ref{Hx}) can  
be a model of decoherence (in which case ${\cal R}$ would be the ensemble of
many degrees of freedom of a ``reservoir'' to which we have only partial
access), or $H(x)$ can generate a unitary evolution if 
${\cal R}$ and ${\cal S}=\{{\cal S}_1,\ldots,{\cal S}_N\}$ are
completely under our control. 
The sum over $\nu$ runs over
an arbitrary number of operators for each subsystem $\cS_i$ and $\clR$, but
$R_\nu$ can also mean 
operators on different subsystems of ${\cal R}$ if ${\cal R}$ is composite
(e.g.~positions 
of harmonic oscillators modelling a heat bath).  In order to have a generic
name for $\clR$ that encompasses these different situations, we will refer
to  $\clR$
 as the ``quantum bus''. The entire 
dependence on $x$ is included in the coupling operators $S_{i,\nu}(x)$.
With ``collective couplings'' (and with ``collective decoherence''
if ${\cal R}$ is a reservoir) we mean $S_{i,\nu}(x)$ which do not depend on
$i$.   

The smallest uncertainty $\delta x$ with which $x$ can be measured is
found from quantum parameter estimation theory 
\cite{Braunstein94}. 
If the state of a system is given by a 
density matrix $\rho(x)$, the smallest achievable $\delta x$ is given by
\begin{equation} \label{dxmin}
\delta x \ge\delta x_{\rm
  min}=\frac{1}{\sqrt{M}\left(\frac{ds^2}{dx^2}\right)^{1/2}}\,, 
\end{equation}
where we allow for
$M$ repetitions of the same measurement in identically prepared states
$\rho(x)$, and $ds^2$ is a metric on the space of density
operators. It is related to the Bures' metric $d_{\rm
  Bures}(\rho,\rho+d\rho)$, with $d\rho=\rho'(x)dx$, by 
$ds=2d_{\rm Bures}(\rho,\rho+d\rho)$. For pure states, the Bures distance
reduces essentially to their overlap,
$d_{\rm
  Bures}(|\psi\rangle\langle\psi|,|\phi\rangle\langle\phi|)=\sqrt{2}\sqrt{1-|\langle  
  \psi|\phi\rangle|}$  \cite{Bengtsson06}.
If $\rho(x)$ and $\rho(x)+d\rho$ are
related through a unitary transformation with generator $\hat{h}$,
$\rho(x+dx)=\exp(-\ri \hat{h}dx)\rho(x)\exp(\ri \hat{h}dx)$, then
\cite{Braunstein94} 
\begin{equation} \label{dbures2}
d_{\rm
  Bures}(\rho(x),\rho(x)+d\rho)=\langle\Delta\hat{h}^2\rangle^{1/2}dx\,.
\end{equation}
An operational definition of $\delta x$ is given by
\begin{equation} \label{dxdef}
\delta x=\frac{\langle \Delta A^2\rangle_x^{1/2}}{\sqrt{M}|\partial \langle A
  \rangle_x/\partial x|}\,, 
\end{equation}
where we see that $\delta x$ corresponds to the quantum uncertainty of
an observable $A$ in state $\rho(x)$, suitably translated by the slope of
$\langle A\rangle_x$ into a fluctuation of $x$.  As usual, for any
observable $A$,  $\langle\Delta A^2\rangle\equiv 
\langle A^2\rangle-\langle A\rangle^2$, and all
expectation values are with respect to $\rho(x)$.  Inequality (\ref{dxmin})
holds for all possible measurements, and for
$M\to\infty$ a measurement exists that saturates the bound \cite{Braunstein94}.

Model (\ref{Hx}) cannot be 
solved in all generality.  However, if the interaction
is sufficiently weak it can be treated in perturbation theory.  Since
we start in an initial product state at $t=0$, we can  then relate properties
of the full model to the single particle dynamics of all $\cS_i$ and $\clR$.  
We first establish the fundamental lower bound on
$\delta x$ with the help of quantum parameter estimation theory, and then
calculate $\delta x$ for a given measurement on $\clR$.  

\subsection*{Quantum parameter estimation theory}
We decompose $H(x)=H_0+H_I(x)$, $H_I(x)=\sum_{i,\nu}S_{i,\nu}(x)\otimes
R_\nu$,
and switch to the interaction picture with respect to $H_0$, with wave
function  $\ket{\psi_I(x,t)}=\exp(\ri H_0
t)\ket{\psi(x,t)}$, $\ket{\psi(x,t)}=\exp(-\ri H(x)t)\ket{\psi_0}$.
In ``Methods'' we show that the Bures distance between the two states
$\rho(x)=\ket{\psi_I(x,t)}\bra{\psi_I(x,t)}$ and
$\rho(x+dx)=\ket{\psi_I(x+dx,t)}\bra{\psi_I(x+dx,t)}$ 
is given by
\begin{eqnarray}
d_{\rm Bures}^2(\rho(x),\rho(x+dx))&=&dx^2\int_0^t\int_0^t dt_1dt_2
K_{|\psi_0\rangle}(H_I'(x,t_1),H_I'(x,t_2)) \label{dbPT}
\end{eqnarray}
where we have defined the correlation function for any two operators $A$,
$B$ in the state $|\psi\rangle$,
$K_{|\psi\rangle}(A,B)=\langle\psi|A B|\psi\rangle-\langle\psi|
A|\psi\rangle\langle \psi|B|\psi\rangle$,
$H_I(x,t)=\exp(\ri H_0t)H_I(x)\exp(-\ri H_0t)$ is the interaction 
hamiltonian in the interaction picture, and $H_I'(x,t)=\partial
H_I(x,t)/\partial x$.   
Equation~(\ref{dbPT}) generalizes (\ref{dbures2}), which is
recovered if $[H'(x),H(x)]=0$ and $H_I(x)t=x \hat{h}$. 
From (\ref{dxmin}), we have
\begin{equation} \label{dxminPET}
\delta x_{\rm min}=\frac{1}{2\sqrt{M}\left(\int_0^t\int_0^t dt_1dt_2
K_{|\psi_0\rangle}(H_I'(x,t_1),H_I'(x,t_2))\right)^{1/2}}\,.
\end{equation}
For identical and identically prepared systems $\cS_i$, $S_{i,\nu}=S_\nu$ and
$|\varphi\rangle_i=|\varphi\rangle$ for
all $i$, and an initial product state 
\begin{equation} \label{inst}
\ket{\psi_0}=\ket{\varphi}^{\otimes
  N}\otimes \ket{\xi}\,,
\end{equation}
 we find   
\begin{eqnarray}
K_{|\psi_0\rangle}(H_I'(x,t_1),H_I'(x,t_2))&=&\sum_{\nu,\mu} \Big(N K_{|\varphi\rangle}(S_\nu'(x,t_1),S_\mu'(x,t_2))\langle
R_\nu(t_1)R_\mu(t_2)\rangle\nonumber\\
&&+N^2\langle S_{\nu}'(x,t_1)\rangle\langle
S_\mu'(x,t_2)\rangle K_{|\xi\rangle}(R_\nu(t_1),R_\mu(t_2))\Big)\,,\label{KCS}
\end{eqnarray}
where the expectation values for operators of $\cS_i$ ($\clR$) are taken in
  states 
  $|\varphi\rangle$ ($|\xi\rangle$). Together with
  eq.~(\ref{dxminPET}) this proves the existence of a measurement on $\cS$ and
$\clR$ that gives
a $1/N$ scaling of $\delta x_{\rm 
  min}$ for $N\gg 1$ and an initial product state, provided
  $\sum_{\nu,\mu}\int_0^t\int_0^t \langle 
  S_{\nu}'(x,t_1)\rangle\langle 
S_\mu'(x,t_2)\rangle K_{|\xi\rangle}(R_\nu(t_1),R_\mu(t_2))dt_1dt_2\ne 0$.

\subsection*{Measuring the quantum bus}\label{sec.mqb}
We now use directly eq.~(\ref{dxdef}) for showing that the $1/N$
scaling can be achieved with the measurement of almost any observable $A$
on $\clR$ alone. The expectation values in eq.~(\ref{dxdef}) are
in general time-dependent. 
This implies a time-dependent minimal uncertainty as well which does,
however,  not affect the scaling with $N$.  We evaluate
$\langle  
A(t)\rangle$ and $\langle\Delta A^2(t)\rangle$ again by using second order
perturbation theory in the interaction.  The general
results for these expressions are cumbersome, but simplify considerably if
we make the following two assumptions: 
(1)~The initial state of $\clR$ is an eigenstate of $A$, 
$A\ket{\xi}=a_\xi\ket{\xi}$; and (2)~$A$ commutes
with $H_R$.  Both assumptions taken together imply that the
quantum 
bus is prepared in a noiseless state at $t=0$ ($\langle \Delta
A^2(0)\rangle=0$). 
Under the above two assumptions, we find 
\begin{eqnarray}
\langle
A(t)\rangle&=&a_\xi+\int_0^tdt_1\int_0^tdt_2\chi_{S\nu\mu}(
N,x,t_1,t_2)C_{RA\nu\mu}(t_1,t_2)+{\cal O}(H_I^3)\,,\label{At}\\  
\chi_{S\nu\mu}(N,x,t_1,t_2)&=&N
K_{|\varphi\rangle}(S_\nu(x,t_1),S_\mu(x,t_2))+N^2\langle
    S_\nu(x,t_1)\rangle\langle S_\mu(x,t_2)\rangle\\
C_{RA\nu\mu}(t_1,t_2)&=&\langle R_\nu(t_1)AR_\mu(t_2)\rangle-a_\xi\langle
R_\nu(t_1)R_\mu(t_2)\rangle \,,
\end{eqnarray}
which can also be used to obtain $\langle
A^2(t)\rangle$ and $\langle
\Delta A^2(t)\rangle$. Eq.(\ref{dxdef}) then leads to
\begin{eqnarray}
\delta x&=&\frac{\Big[
    \int_0^tdt_1\int_0^tdt_2\sum_{\nu,\mu}\chi_{S\nu\mu}(N,x,t_1,t_2)\langle[R_\nu(t_1),A][A,R_\mu(t_2)]\rangle\Big]^{1/2}}{\sqrt{M}\left|\int_0^tdt_1\int_0^tdt_2\sum_{\nu,\mu}\frac{\partial}{\partial 
     x}\chi_{S\nu\mu}(N,x,t_1,t_2)\langle
    R_\nu(t_1)[A,R_\mu(t_2)]\rangle\right|}\,.\label{dxFin}
\end{eqnarray}
In the limit of $N\gg 1$, the term quadratic in $N$ in
    $\chi_{S\nu\mu}(N,x,t_1,t_2)$ dominates, and we find a $1/N$
    scaling of $\delta x$, 
\begin{eqnarray}
\delta x&=&\frac{\Big[
    \int_0^tdt_1\int_0^tdt_2\sum_{\nu,\mu}\langle
    S_\nu(x,t_1)\rangle\langle S_\mu(x,t_2)\rangle\langle[R_\nu(t_1),A][A,R_\mu(t_2)]\rangle\Big]^{1/2}}{\left|\int_0^tdt_1\int_0^tdt_2\sum_{\nu,\mu}\frac{\partial}{\partial 
    x}\left(\langle
    S_\nu(x,t_1)\rangle\langle S_\mu(x,t_2)\rangle\right)\langle
    R_\nu(t_1)[A,R_\mu(t_2)]\rangle\right|}\frac{1}{\sqrt{M}N}\,,\label{dxFin2}
\end{eqnarray}
provided that the denominator does not vanish.  It is enough to measure an 
observable of the quantum bus $\clR$ alone, with all subsystems initially in
a product state.  

\subsection*{Decoherence}\label{sec.dec}
All derivations so far apply perfectly well
if $\clR$ is an environment with many degrees of
freedom which we cannot fully measure.  Measuring an observable $A$ on only
a subset of these implies a non-unitary evolution of
$\cS$. This establishes immediately that we can reach a $1/N$ scaling
of $\delta x$, if $x$ parametrizes a collective decoherence process,
and if we can measure at least some part of the environment. The 
example of superradiance that we will work out below is of this 
type.  However, one might also
be interested in how the unitary
evolution generated by the hamiltonian (\ref{Hx}) is affected by additional
independent 
decoherence of the components $\cS_i$ and $\clR$.  For Markovian
decoherence, such a 
situation is described by a master equation for the density matrix $W(t)$ of
$\cS$ and $\clR$ of the form  
\begin{equation} \label{Wdot}
\dot{W}(t)=-\ri\left(L_0+L_I(x)+\ri\Lambda_s+\ri\Lambda_R\right)W(t),
\end{equation}
where $L_0\,X=[H_0,X]$, $L_I(x)\,X=[H_I(x),X]$, and $\Lambda_R$
($\Lambda_S$) are 
Liouvillians of the Lindblad-Kossakowski type \cite{Breuer06} for $\clR$
($\cS$), with  
$\Lambda_S=\sum_{i=1}^N \Lambda_i$.    

The free evolution ($H_I=0$) 
still factorizes, such that,
essentially, all expectation values
and correlation functions are replaced by expectation values with
respect to the relevant mixed states (see eq.~(\ref{Atd}) in ``Methods'').  
The
$1/N$ scaling is therefore robust 
under individual decoherence of the components, an eventually increased
prefactor not withstanding.  This is corroborated by
further exact results for a pure interaction
with decoherence added to all $\cS_i$ or to $\clR$ (see ``Supplementary
Discussion''), and by the
example of superradiance below.

\subsection*{Measuring the length of a cavity}\label{sec.example} 
As example of an application, we now show how to measure the relative
change of length $\delta 
L/L$ of a cavity with an uncertainty of order $1/N$ with an initial product
state of $N$ quantum resources.  We first consider
unitary evolution.

\begin{figure}[h!t]
\epsfig{file=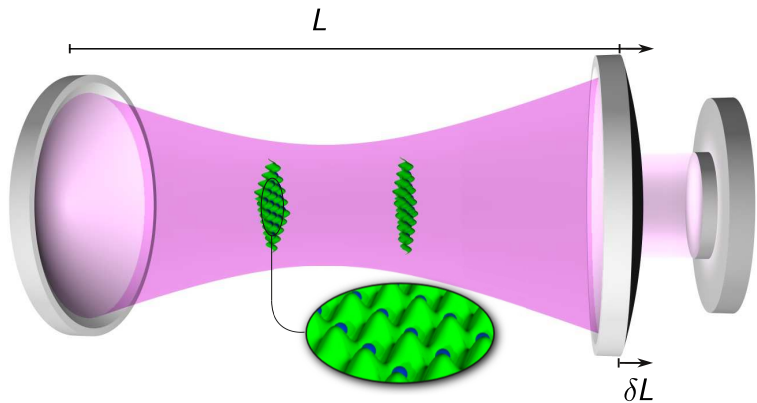,width=10cm,angle=0}\hspace{0.3cm}
\caption{Scheme for measuring the change of the length of an optical
  cavity. $N$ atoms (or ions) are trapped at
  fixed positions in two 2D optical lattices perpendicular to the cavity
  axis. A dipole transition of the atoms is in resonance with a single,
  leaky  cavity mode. The atoms are initially prepared in a dark state in
  which destructive interference prevents 
  the photons from being transferred 
  from the atoms to the cavity mode.  When the cavity length $L$ changes 
  by a small amount $\delta L$, the true dark states evolve, and the initial
  state is exposed to collective 
  decoherence, detectable by photons leaking out through the semi-reflecting
  mirror at a rate proportional to $N^2$. This allows to measure $\delta
  L/L$ with a Heisenberg limited uncertainty of order $1/N$, even if the
  initial dark state is a product state. 
  }\label{fig.system}
\end{figure}

Let $N$ two--level atoms or ions ($N$ even, ground and excited
states $|0\rangle_i$, $|1\rangle_i$ for atom $i$, $i=1,\ldots,N$)
be localized in a cavity, and
resonantly coupled with real coupling constants $g_i$ to a single
e.m.~mode of the cavity of frequency $\omega$ and annihilation operator $a$
(see 
Fig.~\ref{fig.system}),
interaction hamiltonian $H_I=\sum_{i=1}^N g_i\left(\sigma_-^{(i)}a^\dagger
+\sigma_+^{(i)}a\right) 
$,
where $\sigma^{(i)}_-=|0\rangle_{i}\langle 1|_{i}$,
$\sigma^{(i)}_+=|1\rangle_{i}\langle 0|_{i}$.
Due to the spatial dependence
of the e.m.~mode in resonance with the atoms, the $g_i$ depend on
the position $z_i$ of the atoms along the cavity axis and on the
length $L$ of the cavity (the waist of the mode is taken to be much
larger than the size of the atomic ensemble),
\begin{equation} \label{gi}
g_i=\sqrt{\frac{\hbar\omega}{\epsilon_0 V}}\sin(k_z
z_i)\,\beps\cdot\bd\,,
\end{equation}
where $k_z=\pi n_z/L$, $\epsilon_0$  denotes the dielectric constant
of vacuum, $V=LA$ the mode volume (with an effective cross--section
$A$), $\beps$ the polarization vector of the mode, and $\bd$ the
vector of electric dipole transition matrix elements between the
states $|0\rangle_i$ and $|1\rangle_i$, taken identical for all
atoms.

If all $g_i$ are identical, we obtain the Tavis-Cummings model
\cite{Tavis68}. Here we consider the situation where the atoms can be grouped
into {\em 
two sets} with $N/2$ atoms each and coupling constants $G_1$ in the
first set ($i\in \{1\ldots,N/2\}$), and $G_2$ in the second
set ($i\in \{N/2+1,\ldots,N\}$).
One way of obtaining two
coupling constants may be to trap the atoms in two two--dimensional
lattices perpendicular to the cavity  axis (see
Fig.~\ref{fig.system}). Note that it is not necessary to locate
the atoms within a quarter wave-length of each other in order to obtain a
DFS, as would be necessary
without the cavity \cite{PhysRevA.76.012331}.  Distances which are 
integer multiples of the wave-length work just as well.  In (\ref{gi}) we
have 
neglected the transversal dependence of the mode, assuming that the atoms
are localized at a distance from the cavity axis much smaller than the
waist of the mode. However, this is for a computational convenience
only. The initial product DFS states also exist if there is a
radial variation of the $g_i$, but describing the dynamics would become much
more 
complicated as it would depend on all the different $g_i$. 
Assuming two different sets of coupling constants, the system is described
by eq.~(\ref{Hx}), where  
we identify a {\em pair} of atoms $(i,i+N/2)$ with subsystem $\cS_i$,
$i=1,\ldots,N/2\equiv N_p$, and the 
resonant cavity mode with the quantum bus $\clR$.
The free hamiltonian $H_0$ consists of the energy of all atoms,
$H_i=(\omega/2)(\sigma_z^{(i)}+\sigma_z^{(i+N/2)})$, and the energy of the
cavity mode, $H_R=\omega a^\dagger a$. 

An expansion of the $g_i$ about $L$ for a small change
$\delta L$ allows one to write 
the coupling in the form of (\ref{Hx}) with
$
S_{i,1}(x)=g\left((1+x)\sigma_-^{(i)}+(1-x)\sigma_-^{(i+N/2)}\right)
$,
$S_{i,2}=S_{i,1}^\dagger$, and $R_1=a^\dagger$, $R_2=a$, and $x\propto
\delta L/L$ (see 
SI for the prefactor).
For notational simplicity we restrict ourselves to  
the case where for $x=0$ the couplings are the same for the
two sets, $G_1=G_2=g$, but this is by no means necessary
for the method to work.

A convenient basis for a pair of atoms is given by the ``singlet'' and
``triplet'' states $\{|s\rangle,\ket{t_-},\ket{t_0},\ket{t_+}\}$ with
$\ket{s}=(\ket{01}-\ket{10})/\sqrt{2}$, $\ket{t_-}=\ket{00}$,
$\ket{t_0}=(\ket{01}+\ket{10})/\sqrt{2}$, and $\ket{t_+}=\ket{11}$.  As initial
state of all $\cS_i$ and $\clR$ we take the product state (\ref{inst}) with
$N\to N/2$, and
$\ket{\varphi}=(\ket{t_-}+\ket{s})/\sqrt{2}$, and $|\xi\rangle=\ket{0}$ for
a cavity mode in the vacuum state. We obtain a time-independent 
$K(t_1,t_2)=g^2\left(N_p+N_p^2\right)/2$, and from (\ref{dxminPET}) 
\begin{equation} \label{dxSR}
\delta x_{\rm
  min}=\frac{\sqrt{2}}{\sqrt{M}gt\sqrt{2N+N^2}}\,, 
\end{equation}
which clearly scales as $1/N$ for $N\gg 1$. One might argue that a small
amount of entanglement is present in $\ket{\varphi}$, but the size of
the cluster of atoms all entangled with each other (i.e.~a pair of atoms) is
independent of $N$, 
such that it is legitimate to consider a pair of atoms as individual
subsystem, and it is a product state of these subsystems that we consider.
In  SI we show that the product state can be prepared 
by letting the atoms interact pairwise.    

The initial state contains half a
photon per atom.  For a generic  state the excitations stored in the
atoms would start oscillating between 
the cavity mode and the atoms.  However, for $x=0$ our initial state is a
``dark state'', as destructive interference prevents the transfer of the
photon from any pair of atoms to the cavity.  When $x$ deviates from
zero, the perfect cancellation in the destructive interference is broken,
and photons get transferred to the cavity.  

Measuring the
number of photons constitutes an optimal measurement in the sense that the
bound (\ref{dxSR}) is reached. 
To see this, we identify $A=a^\dagger a$ in eq.~(\ref{dxFin2}). This leads
in a straightforward manner to 
$\delta x=\frac{\sqrt{2}}{\sqrt{M}g t\,N}$,
which agrees with (\ref{dxSR}) for $N\gg 1$, including the
prefactor.
After what was said in Sec.~``Decoherence'', it is clear that adding
independent decoherence to all subsystems does not change the $1/N$
scaling of $\delta x_{\rm min}$.  We now show this
explicitly by considering the situation of very strong damping of the
cavity mode, the superradiant regime.

The framework of Sec.~``Decoherence''
is suited for this analysis, but 
we adopt the well-developed theory of superradiance 
\cite{Agarwal70,Bonifacio71a,Glauber76,Gross82} to give an independent
demonstration that $\delta x_{\rm min}$ scales as $1/N$. Decoherence arises
because of two processes: 
Each atom can undergo 
spontaneous emission with rate $\Gamma$, due to its coupling
to a continuum of additional e.m.~modes. 
The damping of the cavity mode arises from the escape of photons  with a rate
$2\kappa$ through one of the mirrors. 
In the notation of eq.(\ref{Wdot}), and identification of a pair of atoms 
$(i,i+N/2)$ with $\cS_i$, 
the generators $\Lambda_i$ and $\Lambda_R$ for these two processes read
\cite{Agarwal70,Bonifacio71a,Glauber76,Gross82}
\begin{eqnarray} \label{LiX}
\Lambda_i X&=&  \frac{\Gamma}{2}\left([\sigma_-^{(i)}
  X,\sigma_+^{(i)}]+[\sigma_-^{(i+N/2)}
  X,\sigma_+^{(i+N/2)}]+h.c.\right)\,, \\
\Lambda_R X&=&\kappa\left([aX,a^\dagger]+h.c.\right)\,.\label{LRX}
\end{eqnarray}
Superradiance occurs in the overdamped regime $\Gamma\ll g\sqrt{N}\ll
\kappa$, where a photon transferred to the cavity 
leaves the cavity before it can feed itself back to the atoms, but induces
emission in other atoms while in the cavity mode. Cavity decay is then the
by far dominant process.  We will therefore start by neglecting $\Gamma$,
but treat spontaneous emission in SI. The
population of the cavity follows the occupation of the atoms adiabatically,
and one can eliminate the cavity mode.  This leads to the 
well-known and, for $x=0$, experimentally 
verified master equation of
superradiance
\cite{Agarwal70,Bonifacio71a,Glauber76,Gross76,Skribanowitz73,Gross82}
for the reduced density matrix $\rho_s$ of the atoms in the interaction
picture, 
\begin{equation} \label{L}
\frac{d}{dt}\rho_s(t)=L_I(x)[\rho_s(t)]\equiv \gamma\,([J_-(x)\rho_s(t),J_+(x)]+[J_-(x),\rho_s(t) J_+(x)])\,.
\end{equation}
The collective
generators $J_\pm$ are
$J_-(x)=\sum_{i=1}^{N/2}S_{i,1}(x)=\sum_{i=1}^{N/2}\left((1+x)\sigma_-^{(i)}+(1-x)\sigma_-^{(i+N/2)}\right)$, 
$J_+(x)=J_-^\dagger(x)$.  The rate
$\gamma=g^2/\kappa$ is independent of $N$. 
Collective decoherence is a two-stage process here, as photons
stored in the atoms first need to be transferred to the cavity mode before
they can leave the system.  The dark states of Sec.~``Measuring the length
of a cavity'' 
are therefore 
decoherence-free states. There is a large decoherence-free subspace (DFS)
containing ${N \choose N/2} \sim 2^{N}/\sqrt{N}$ DF
states, including a 
$2^{N/2}$ dimensional subspace
$\bigotimes_{l=1}^{N/2}\{|t_-\rangle_l,|s\rangle_l\}$ in which the
pair formed by the atoms $l$ and $l+N/2$  can be in a superposition
of $|t_-\rangle_l$ and 
$|s\rangle_l$ \cite{Beige00b,Braun01B}. A DFS of the same
dimension also exists for non-identical
couplings, but the coefficients in the
linear combination of the singlet state need to be adapted accordingly,
$|s\rangle_l
\to\frac{1}{\sqrt{|{G_1}|^2
    +|{G_2}|^2}}({G_1}|0\rangle_l|1\rangle_{l+N/2}-{G_2}    
  |1\rangle_l|0\rangle_{l+N/2})$. 
If after preparing the atoms in a DFS state corresponding
to the initial couplings ${G}_1^{(0)}$, $G_2^{(0)}$
the length $L$ of the cavity changes slightly, the coupling
constants will evolve,  ${G}_I^{(0)}\rightarrow
{G}_I$, $I=1,2$, and  so will the DFS.
Photons will leak out of the cavity as the original
state becomes exposed to decoherence. 

There is a well-known connection between the photon statistics in the
 cavity mode and the excitation of the atoms, derived in \cite{Bonifacio71a}
 for 
all couplings identical,
\begin{equation} \label{aJ}
\langle a^{\dagger
  m}a^m(t)\rangle=2m\left(\frac{g}{\kappa}\right)^{2m}\int_0^t\,ds \,\kappa\,
  e^{-2m\kappa s}(e^{\kappa s}-1)^{2m-1}\langle 
J_+^m J_-^m(t-s)\rangle\,.
\end{equation}
One checks that this relation remains valid for small asymmetries
 $x\ne 0$.  Thus, instead of the
 number of photons in the cavity for a given value $x$ one can calculate
 the excitation of the 
 atoms, where, however, the observable itself becomes a function of $x$,
 $J_+^m(x) J_-^m(x)$.  For the initial product state (\ref{inst}) considered above (with
$N\to N_p$, 
$\ket{\varphi}=(\ket{t_-}+\ket{s})/\sqrt{2}$, and $|\xi\rangle=\ket{0}$)
 we have from (\ref{L}), 
\begin{eqnarray}
\langle J_+J_-(t)\rangle&=& x^2\left(\frac{N_p}{2}+\frac{N_p^2}{2}\right)-\gamma t \left(
3N_p^2+N_p^3\right)(x^2+x^4)+{\cal O}(t^2)\,,\label{jpjm}\\
\langle
J_+^2J_-^2(t)\rangle&=&\left(\frac{N_p}{2}-\frac{5N_p^2}{4}+\frac{N_p^3}{2}+
\frac{N_p^4}{4}\right)x^4
\nonumber\\
&+&2\gamma
t\bigg[\left(-N_p-\frac{5}{2}N_p^2+\frac{11}{2}N_p^3-\frac{3}{2}N_p^4-\frac{1}{2}N_p^5\right)x^4
  \nonumber\\
&+&\left(-3N_p+\frac{5}{2}N_p^2+\frac{7}{2}N_p^3-\frac{5}{2}N_p^4-\frac{1}{2}N_p^5
\right)x^6
  \bigg]+{\cal O}(t^2)\,. \label{jp2jm2}
\end{eqnarray}
Equation~(\ref{aJ}) is in principle valid only in the
 Markovian regime $t\gg 1/\kappa$, if $\langle 
J_+^m J_-^m(t-s)\rangle$ is obtained from the solution of the Markovian
 superradiance master equation (\ref{L}).  However, the initial behavior of 
 $\langle a^{\dagger } a (t)\rangle$, $\langle a^{\dagger  } a
 (t)\rangle\simeq g^2t^2\langle J_+J_-(0)\rangle$, is entirely determined by
 the value of $\langle  
J_+ J_-(t)\rangle$  at $t=0$, i.e.~the question of the Markovian
 approximation of the dynamics of $\langle 
J_+ J_-(t)\rangle$ does not arise, and eq.~(\ref{aJ}) can therefore be used
 to calculate $\langle n_{\rm ph}(x,t)\rangle$ for short times up to order
 $t^2$.  From $\langle J_+J_-(0)\rangle$ one finds immediately 
\begin{equation} \label{nphSR}
\langle n_{\rm ph}(x,t)\rangle=\langle a^\dagger
a(x,t)\rangle=\frac{1}{2}g^2t^2x^2\left(N_p(N_p+1)\right)+{\cal O}(t^3)\,.
\end{equation}
At this order
$\kappa$ does not intervene yet, as initially the cavity mode
is in the vacuum state. The quadratic initial increase of $\langle n_{\rm
  ph}\rangle$ reflects the beginning of a Rabi oscillation between
the excited atoms and the cavity mode. We expect this result therefore to be
valid as long as $\langle n_{\rm ph}(x,t)\rangle\lesssim 1$. 
Equation~(\ref{nphSR}) agrees identically with the result one finds from the
approach in Sec.~``Decoherence'' (see
eq.(\ref{Atd}) in ``Methods''). 
The fluctuations of $n_{\rm ph}$ are obtained from
$\langle n_{\rm
  ph}(x,t)^2\rangle  = \langle a^{\dagger 2}a^2(x,t)\rangle+\langle n_{\rm
  ph}(x,t)\rangle$. Together with (\ref{aJ}) one gets for 
  $\langle n_{\rm ph}(x,t)\rangle\lesssim 1$, 
$\langle \Delta n_{\rm ph}^2(x,t)\rangle\simeq\langle n_{\rm
  ph}(x,t)\rangle$
with corrections of order $(g/\kappa)^4$. From
eqs.(\ref{dxdef}) and (\ref{nphSR}) we find  
$\delta
x=\frac{\sqrt{2}}{\sqrt{M}gt\sqrt{N(N+2)}}\simeq\frac{\sqrt{2}}{\sqrt{M}
  gtN} $,
which is identical to the minimal possible uncertainty, eq.(\ref{dxSR}) 
for $N\gg 1$.  
The validity of the short time expansions
(\ref{jpjm},\ref{jp2jm2}) is limited to $N\gamma t\ll 1$, as can be seen
from comparing the first order term with the zeroth order term.  Inserting
(\ref{jpjm}) in  (\ref{aJ}) gives therefore an analytical prediction 
of
$\langle n_{\rm 
  ph}(t)\rangle$ valid for $g/\kappa\ll gt\ll \kappa/(Ng)$, in addition to the
small time result (\ref{nphSR}) for $gt\ll g/\kappa$. The agreement of
$\langle n_{\rm 
  ph}(t)\rangle$ based on (\ref{jpjm}) with the result from simulating
(\ref{L}) can  be further improved by re-exponentiating $\langle
J_+J_-(t)\rangle$ according to $a+bt\simeq a\exp(b/a t)$, before inserting
it in (\ref{aJ}). The limitation of validity of the small-time expansion
does not pose a serious restriction in the bad cavity limit $\kappa\gg g$,
nor does it imply that the $1/N$ scaling of the sensitivity breaks down
beyond that regime. A full theoretical analysis for longer times will have
to include the calculation of the superradiant propagator with broken
$SU(2)$ symmetry, however.
For $t\simeq 
1/\kappa$ a  
non-Markovian description of superradiance is called for, which is beyond
the scope of the present investigation.  

\begin{figure}[h!t]
\epsfig{file=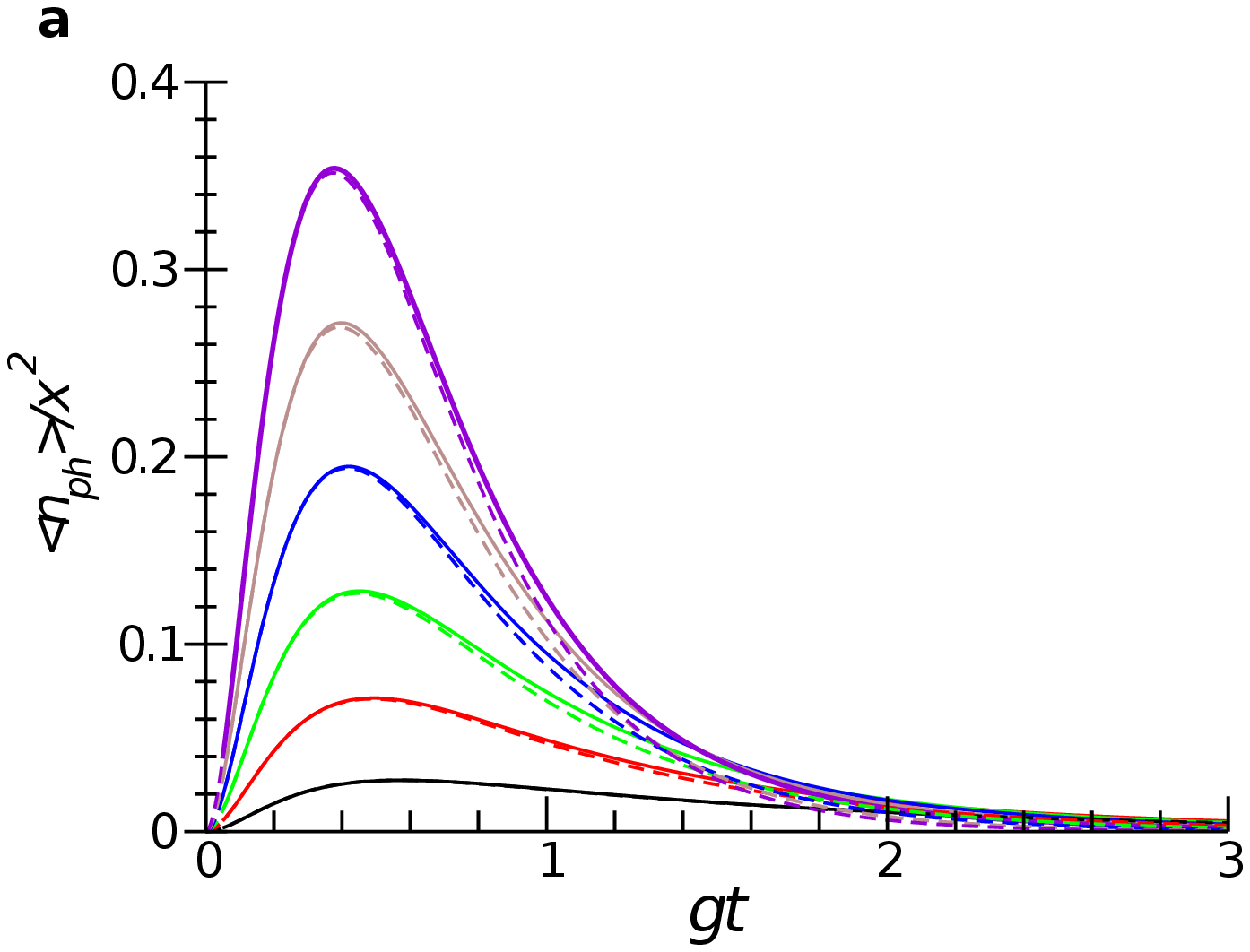,width=8cm,angle=0}\hspace{0.25cm}
\epsfig{file=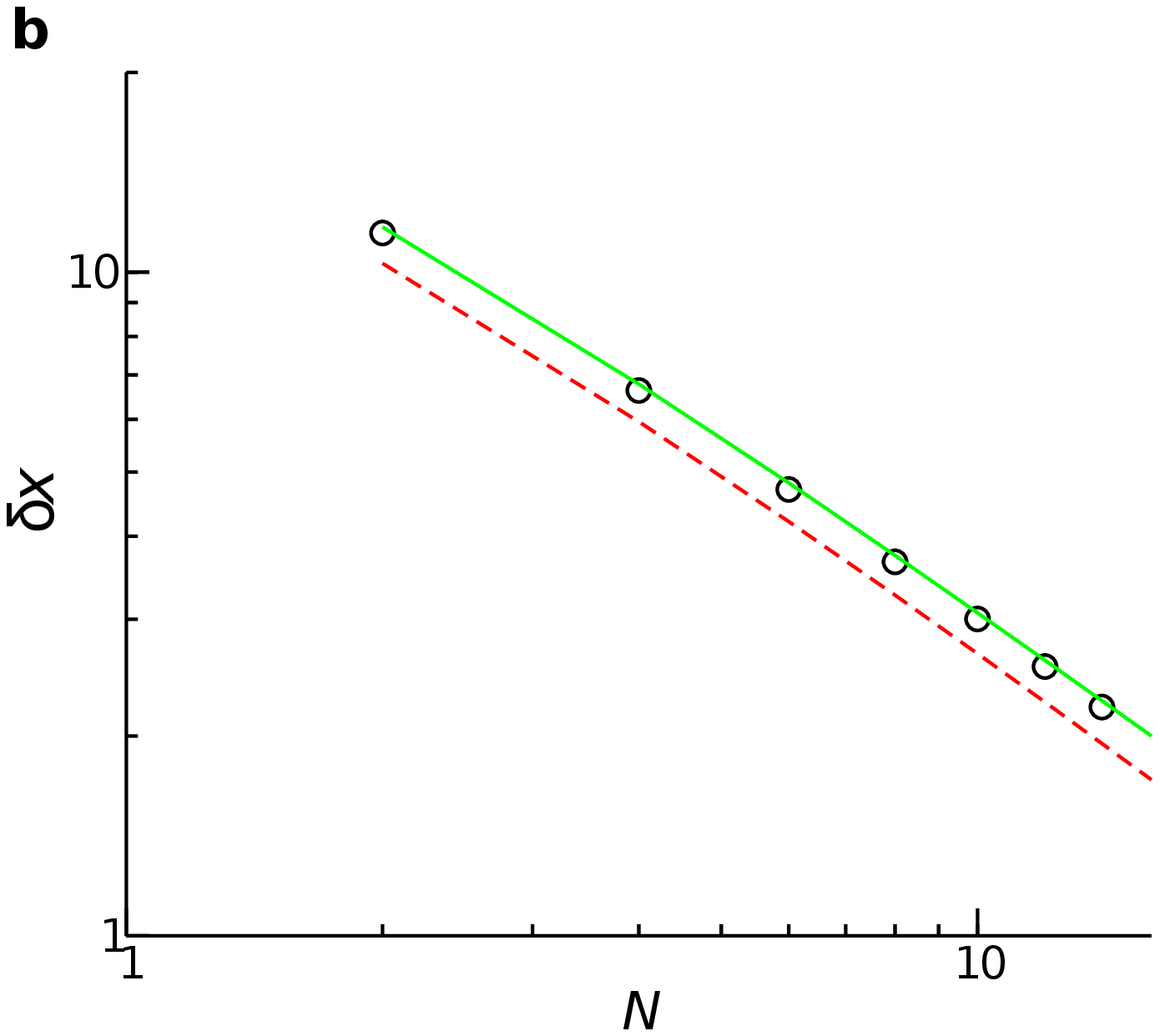,width=7cm,angle=0}
\caption{Mean photon number
  and uncertainty $\delta x$ of the change of length of the cavity {\bf (a)}:
  Mean photon number $\langle 
  n_{\rm ph}\rangle$ as a function of  
  dimensionless time $gt$ (where $g$ is the coupling constant of the atoms
  to the cavity mode) for $N=2,4,6,8,10,12$ (black, red, green, 
  blue, brown, violet), in units of $x^2$ for $x=0.1$, and with photon escape
  rate from cavity $\kappa=5g$, obtained
  through 
  numerical simulation of superradiance using a stochastic Schr\"odinger
  equation.   The 
  dashed lines with corresponding colors are analytical results valid up to
 $Ng^2t/\kappa\sim 1$ (see eqs.~(\ref{aJ}),(\ref{jpjm})).  {\bf (b)}:
  Uncertainty $\delta x$ 
  (see eq.~(\ref{dxdef})) based on $A=n_{\rm ph}$ as function of  
  $N$ for $gt=0.0485$ and $x=0.01$.  Numerical results
  (circles)   show the same $1/N$ scaling as the ideal lower bound
  (red dashed line),
  eq.~(\ref{dxSR}), with slightly increased 
  prefactor. Green continuous line is an 
  analytical prediction based on an expansion of $\langle n_{
\rm ph}(t)\rangle$ for small $gt$.}\label{fig.nph} \vspace{0.5cm}
\end{figure}

Figure \ref{fig.nph} shows that $\langle n_{\rm
  ph}(t)\rangle$ obtained numerically by simulating (\ref{L})  
through an equivalent stochastic Schr\"odinger equation (SSE),
and integration of $\langle J_+J_-(t)\rangle$ according to eq.~(\ref{aJ}),
agrees 
well with the result based on (\ref{jpjm}) for $gt\ll g/\kappa$ and
$g/\kappa\ll gt\ll \kappa/(Ng)$.  The SSE for real $\psi(t)$ reads 
\begin{equation}
d\psi(t)=D_1(\psi(t))\,dt+D_2(\psi(t))\,dW(t)\,,\label{SSE}
\end{equation} 
with
  $D_1(\psi)=\gamma\left(2\langle J_-\rangle_\psi J_--J_+J_--\langle
  J_-\rangle^2_\psi\right)\psi$, and
$D_2(\psi)=\sqrt{2\gamma}\left(J_--\langle J_-\rangle_\psi\right)\psi$, 
where $dW(t)$ is a Wiener process with average zero and variance
$dt$, and $\langle J_-\rangle_\psi=\langle \psi|J_-|\psi\rangle$
\cite{Breuer06}. We used 2000 equidistant time steps in the time interval
$t=0,\ldots,20/g$, 20 random realizations of the process for the simulation
of $\langle n_{\rm ph}(t)\rangle$, and 400 realizations for the calculation of
$\delta x$.  
Figure \ref{fig.nph} also shows $\delta x$ calculated from the numerical data
for $\langle \Delta n_{\rm
  ph}^2(t)\rangle^{1/2}$ and $\langle n_{\rm ph}(x,t)\rangle$
through eq.(\ref{dxdef}), together with the fundamental 
lower bound $\delta x_{\rm min}$, eq.(\ref{dxSR}). We see that at $gt=0.0485$,
$\delta x$ follows the optimal $1/N$ scaling with only slightly 
increased prefactor. 

We emphasize that $n_{\rm ph}$ allows to measure $\delta L/L$, not just
to detect a change of $L$. Eqs.(\ref{aJ}) and (\ref{jpjm}) relate $\langle
n_{\rm ph}\rangle$ to $x$, and, unless the two lattices are situated at
anti-nodes of the mode, the relation between $\delta G\equiv (G_1-G_2)/2=
g x$ and 
$\delta L/L$ is linear to lowest order and independent of $N$:  If we choose
the position 
of the atoms such that $z_2-z_1=m \lambda$  with $n_z-1\ge m\in
\mathbb{N}$ we have
\begin{equation*}
x=\frac{\delta {G}}{g}=m\pi \cot\left(\frac{n_z\pi
z_1}{L}\right)\frac{\delta L}{L}.
\end{equation*}
Therefore, the measurement of $\langle
n_{\rm ph}\rangle$ 
allows the measurement of $\delta L/L$.
Several other practical questions, e.g.~the preparation of the
initial state, and the robustness 
of the method with respect to fluctuations of the coupling constants,
spontaneous emission, and errors in the preparation
of the initial state, are addressed in SI.
The superradiant regime has the advantage of providing direct
access to the 
number of photons in the cavity. The average number of photons outside
is simply obtained by 
integrating $2\kappa\langle a^\dagger a(t)\rangle$ up to time $t$, as the
photon escape  
rate is proportional to the average photon number inside the cavity
\cite{Bonifacio71a}. 
The results for the scaling of $\delta x$ with $N$ based on $A=a^\dagger a$
are therefore unaffected by detecting the photons that leave the
cavity. Measuring the number of
photons amounts 
to monitoring the decoherence dynamics, and we have thus
an example where the parametric dependence of a collective decoherence
process allows to achieve the Heisenberg limit with an initial product state.

\section*{\large Discussion}
Our results may seem to conflict with the well-known theorem
\cite{Giovannetti06} that for unitary evolution of $N$ 
independent quantum 
systems in an initial product state at best a scaling $\delta x_{\rm
  min}\sim 1/\sqrt{N}$  is possible.  To see that there is no contradiction,
it is helpful to consider the simple case where $H'(x)\equiv 
dH(x)/dx$ and $H(x)$ commute, 
$[H(x),H'(x)]=0$.   One then easily
shows that to  
lowest order in $dx$, $\rho(x,t)=\exp(-\ri H(x)t)\rho(0)\exp(\ri H(x)t)$ (with
$\hbar=1$)  and $\rho(x,t)+d\rho$ are related by a
unitary transformation with generator $\hat{h}=H'(x)\,t$. Let us furthermore
restrict ourselves to a   
single operator per subsystem, i.e.~$\nu=1$ only, and to the linear $x$
dependence $S_{i,1}(x)=x S$ for all $i$, and $R_1=R$.
A few lines of calculation lead to 
\begin{equation} \label{dh2}
\langle \Delta\hat{h}^2\rangle=\left(N\langle \Delta S^2\rangle\langle
R^2\rangle +N^2\langle S\rangle^2\langle \Delta R^2\rangle\right)t^2 \,,
\end{equation}
for an initial 
product state, $\rho(0)=\ket{\psi_0}\bra{\psi_0}$ with $\ket{\psi_0}$ from
eq.(\ref{inst}). 
All expectation values of $S$ in (\ref{dh2}) are in state 
$\ket{\varphi}$, those of $R$ in the state $\ket{\xi}$.  Inserting
(\ref{dh2}) into (\ref{dbures2}) and (\ref{dxmin}), we find that for $N\gg 1$
and $\langle S\rangle^2\langle \Delta R^2\rangle\ne 0$,
\begin{equation*}
\delta x_{\rm min}=\frac{1}{2\sqrt{M}t|\langle S\rangle|\langle\Delta,
  R^2\rangle^{1/2}}\frac{1}{N}
\end{equation*}
i.e.~the
Heisenberg limit $\delta x_{\rm min}\sim 1/N$ can be achieved with an
initial product state.  
Clearly, for the case considered above the unitary
transformation generated by $H_0\equiv\sum_i H_i+H_R$ is not
necessary in order to achieve the $1/N$ 
scaling. We therefore simplify the reasoning further  by
considering the case $H_0=0$.   We are then left with a pure 
interaction,
\begin{equation} \label{HI}
H(x)=H_I(x)=x \sum_i S_i\otimes R\,.
\end{equation}
But this is {\em not} a hamiltonian of the form
$H(x)=x\sum_i h_i$ required by the theorem in \cite{Giovannetti06}. In our
case all  
subsystems couple in a non-trivial fashion to the common quantum bus ${\cal
  R}$ and are therefore not independent.
This turns out to be the 
decisive difference.  The SQL can be recovered for the standard situation of
$N$ {\em independent} 
subsystems through a  
$R$ that acts only trivially on ${\cal R}$, i.e.~$R=\mathbf{1}$, such that
$\langle \Delta R^2\rangle=0$, and thus $\delta x_{\rm
  min}= 
1/(2\sqrt{MN}t\langle \Delta S^2\rangle^{1/2})$. This makes obvious the
rather ironic fact that 
quantum 
fluctuations in $\clR$ help and are necessary to achieve the $1/N$
scaling.  The prefactor of the $1/N$ behavior is smallest for an initial
state with an equal weight superposition of the eigenstates of $R$
pertaining to its largest and lowest
eigenvalues $r_{\rm min}$ and $r_{\rm max}$,  in  which case $\langle\Delta
R^2\rangle^{1/2}=|r_{\rm max}-r_{\rm min}|/2$. 

\begin{figure}[h!t]
\epsfig{file=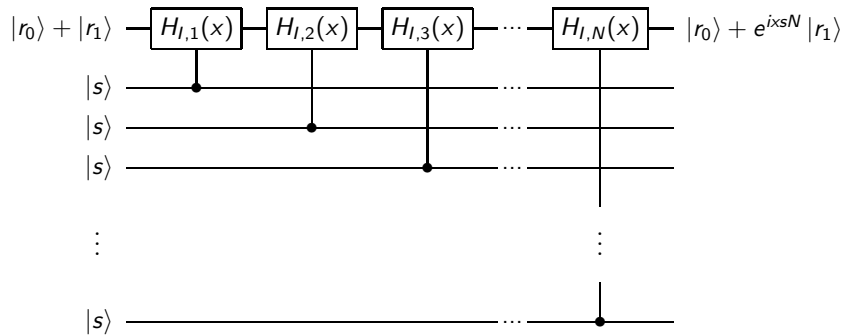,width=11cm,angle=0}\hspace{0.3cm}
\caption{Quantum circuit that reproduces the interaction hamiltonian
  $H_I(x)=x \sum_{i=1}^N S_i\otimes R$. $N$ quantum systems $\cS_i$ prepared
  all in the 
  eigenstate 
  $\ket{s}$ of $S_i$, $S_i\ket{s}=s\ket{s}$ in eq.~(\ref{HI}), lead to a total
  accumulated  
  relative phase between states of the quantum bus $\clR$ that is
  proportional to 
  $xN$.  This allows a measurement of $x$ with a precision that scales as
  $1/N$,   even though the initial state is a product state, by 
  measuring any observable $A$ on $\clR$ alone that does not commute with
  $R$. }
\label{fig.qc}  
\end{figure}

This simple example also allows to corroborate that in order to achieve the
$1/N$  
scaling one need not measure the $\cS_i$ {\em at all}, and almost any
measurement on ${\clR }$ suffices. 
Consider an initial
product state
$|\psi_0\rangle=\otimes_{i=1}^N\ket{s_i}_i|\xi\rangle$, with
$S_i|s_i\rangle=s_i\ket{s_i}$, and 
$\ket{\xi}=\sum_m
d_m|r_m\rangle$  for $R|r_m\rangle=r_m\ket{r_m}$. We then have
\begin{equation} \label{psitH0}
\ket{\psi(t)}=\sum_m d_m e^{-\ri x \sum_i s_i r_m
  t}\bigotimes_{i=1}^N\ket{s_i}\otimes\ket{r_m}\,. 
\end{equation}
Let $A$ be an observable on $\clR$ which does not commute with $R$,
i.e.~there are at least two eigenstates $\ket{r_0}$ and $\ket{r_1}$ such
that $\bra{r_0}A\ket{r_1}\ne 0$. It is sufficient to consider an
initial state of $\clR$ which is a superposition of these two states, e.g.~we
may take $d_0=d_1=1/\sqrt{2}$, and an observable
$A=\ket{r_0}\bra{r_1}+\ket{r_1}\bra{r_0}$. If all subsystems $\cS_i$ are
prepared in the same state with $s_i=s$, one finds
$\langle A(t)\rangle=\cos(x N s(r_0-r_1)t)$ and $\langle\Delta
A^2(t)\rangle=\sin^2(x Ns (r_0-r_1)t)$. Inserted in
eq.~(\ref{dxdef}) this leads to the exact result
\begin{equation} \label{dxsm}
\delta x=\frac{1}{N|s| |r_0-r_1|t}\,,
\end{equation}
valid for all $x$.
This shows that the $1/N$ scaling can be reached by
measuring 
almost any observable of $\clR$, as long as it does not commute with
$R$. Furthermore, 
eq.~(\ref{psitH0}) allows a simple quantum information theoretical
explanation of the effect:  The final state reflects the accumulated phase
from the interaction of all the systems
$\cS_i$ with the common quantum bus $\clR$. Figure~\ref{fig.qc} shows an
equivalent quantum circuit that reproduces 
state (\ref{psitH0}).  One subsystem $\cS_i$ after
another imprints the same phase on the components of the state of
$\clR$.  
Equation~(\ref{psitH0})
also makes obvious that a measurement of
the $\cS_i$ alone does not allow to achieve the $1/N$ scaling, as the state
of $\clR$ will collapse on  a single state $|r_m\rangle$, and one only gets an
irrelevant global phase. Thus, measuring the quantum bus is not only
sufficient, but also 
necessary for the $1/N$ scaling of $\delta x$. We also see that the
measurement of 
$A=\ket{r_0}\bra{r_1}+\ket{r_1}\bra{r_0}$ is optimal if $r_0,r_1$ correspond
to the smallest and largest eigenvalues of $R$, respectively.

In \cite{Higgins07} an adaptive measurement technique was demonstrated that
allows one to achieve Heisenberg-limited uncertainty by using only an
initial product state. The method is based on phase estimation
\cite{Kitaev96}, but instead of using a NOON state of $N$ photons, independent 
photons were passed $N$ times through the same phase shifter.  This
amplifies the phase by a factor $N$, but it was shown that in the presence
of losses the scaling of the sensitivity with $N$ is at most improved by a
constant factor 
\cite{Kolodynski10} compared to the classical case for $N\to\infty$. A
common  feature of both phase estimation and our method 
is that a measurement is performed on a common quantum system that interacts
with all other quantum systems.  However, our method is more
general. It incorporates decoherent and unitary evolutions in the
same framework, and  allows one to use
collective decoherence as a signal. Secondly, phase estimation was developed
for a multi-qubit 
system with controlled, sequentially turned on interactions, and an
$x$-dependence in the free evolution. 
Hamiltonian (\ref{Hx}) on the other hand can be used to describe
substantially more
complex systems, with possibly non-trivial dynamics in the absence of
the collective interaction, and with interactions that do not commute with the
Hamiltonians of the free constituents.  Furthermore, the interaction is
simultaneous such there is no bandwidth penalty in the accumulation of the
phase, nor is there a need to re-sample a phase shift many times.

Our method requires a collective interaction between $N$ separable quantum
systems and a 
common quantum bus, an initial noise-less state in the sense discussed
above, and the possibility
to measure at least part of the quantum bus.  In the case of
incomplete measurement of the 
quantum bus this implies the need of a
collective decoherence process with a decoherence-free initial state.
Besides atoms in a cavity one might consider circuit-QED 
systems \cite{Fink09}, trapped ions coupled to a common
phonon mode \cite{Haeffner08}, or 
quantum dots coupled to micro-resonators \cite{reithmaier_strong_2004} or
to photonic crystals \cite{Ganesh07}. Both 
unitary evolution or a decoherence process can be useful, as long as
the collective interaction 
between the $N$ quantum resources and a common quantum bus 
depends on the parameter $x$ to be measured. 

To summarise, we have developed a general theory of collectively enhanced
quantum measurements based on the interaction of $N$ quantum
systems with a 
common ``quantum bus''.  The latter can be a simple quantum system, or an
environment with many degrees of freedom to which we have only partial
access.   We have shown 
that if the collective interactions depend on a parameter $x$, the
Heisenberg limit (i.e.~a $1/N$
scaling of the uncertainty of $x$) can be reached with  
an initial product state, and by measuring almost any observable of the
quantum bus.  
We have used quantum parameter estimation theory to establish that 
a  $1/N$ scaling of the uncertainty
is indeed optimal in this setup.  We have given a simple quantum-information
theoretical 
interpretation of the effect,  and we have
analysed in detail a possible experimental 
implementation of the measurement of the change of the length of a
cavity with an uncertainty that scales as $1/N$.

The proposed measurement principle offers an attractive way out of the
dilemma of 
ubiquitous decoherence that has so far plagued quantum enhanced
measurements: First of all, there is no need to build highly entangled
states which are extremely fragile under decoherence for large $N$.  Simple
product states will do, and decoherence of some parts of the system does not
affect the $1/N$ scaling of the minimal uncertainty.  Secondly, 
parameter-dependent collective decoherence is covered itself by our new
measurement principle.  Indeed, decoherence is a process in which
quantum 
interference effects can play an important role. This is exemplified by
the very existence of DFS,  and can lead to exquisite sensitivity when
a DFS is disturbed.
Instead of trying to suppress decoherence at
all costs, one might therefore be better off exploiting its parametric
dependence. 

\section*{\large Methods}
\subsection*{Bures distance for unitary evolution}
The state vector 
$\ket{\psi_I(x,t)}$ in the interaction picture obeys the time-dependent
Schr\"odinger equation 
\begin{equation} \label{SEPT}
\ri\frac{\partial}{\partial t}\ket{\psi_I(x,t)}=H_I(x,t)\ket{\psi_I(x,t)}\,,
\end{equation}
with the interaction hamiltonian 
\begin{eqnarray} \label{HIx}
H_I(x,t)&\equiv&\sum_{i,\nu}S_{i,\nu}(x,t)\otimes R_\nu(t)\,,\\
S_{i,\nu}(x,t)&=&e^{\ri H_i t}S_{i,\nu}(x)e^{-\ri H_i
  t},\,\,\,\,R_\nu(t)=e^{\ri H_R 
  t}R_{\nu}e^{-\ri H_R t}\,. \label{srt}
\end{eqnarray}
The general solution of (\ref{SEPT}) is given by 
$\ket{\psi_I(x,t)}={\rm T}
\exp\left[-\ri\int_0^tH_I(x,t')dt'\right]\ket{\psi_0}
$,
where ${\rm T}$ denotes the time-ordering operator.  To second order in the
perturbation $H_I(x,t)$, the overlap between $\ket{\psi_I(x,t)}$ and
$\ket{\psi_I(x+dx,t)}$ reads 
\begin{eqnarray}
\langle
\psi_I(x+dx,t)|\psi_I(x,t)\rangle&=&1+dx\Big[\ri\int_0^t\langle H_I'(x,t_1)
\rangle dt_1+\int_0^t\int_0^{t_1} \langle[H_I'(x,t_1),H_I(x,t_2)]\rangle
dt_1dt_2 \Big]
\nonumber\\
&&+\frac{1}{2}dx^2\Big( \ri\int_0^t\langle H_I^{''}(x,t_1)\rangle dt_1
+\int_0^t\int_0^{t_1}
\langle [H_I^{''}(x,t_1),H_I(x,t_2)]\rangle dt_1dt_2\nonumber \\
&&-2\int_0^t\int_0^{t_1}\langle
H_I'(x,t_1)H_I'(x,t_2)\rangle dt_1dt_2\Big)
\,,
\end{eqnarray}
with all expectation values with respect to $\ket{\psi_0}$.
We assume that the derivatives of $H_I(x)$ with respect to $x$ are hermitian
operators, in which case the term linear 
in $dx$ is purely imaginary.  The lowest order term in the squared
overlap is then of order $dx^2$.  One finds in a straightforward manner
the squared Bures distance (\ref{dbPT}).

\subsection*{Decoherence of subsystems}\label{sec.MethDecAllS}
Markovian decoherence of the $\cS_i$ and $\clR$ on top of the unitary
evolution generated by $H(x)$ can be described by
eq.~(\ref{Wdot}). The single system
dynamics (i.e.~all $\cS_i$ and $\clR$ taken 
separately, $L_I=0$), can be
solved formally by
exponentiating the Liouvillians.   We
will again treat $L_I(x)$ in perturbation theory. The density matrix
$W_I(t)$ of $\cS$ and $\clR$ in the 
interaction picture is related to the one in the Schr\"odinger picture,
$W(t)$, by 
\begin{equation} \label{WI}
W(t)=e^{-\ri (L_0+\ri \Lambda_s+\ri\Lambda_r)t}W_I(t)\equiv e^{-\ri K t}W_I(t)\,,
\end{equation}
and obeys the master equation
$\dot{W}_I(t)=-\ri L_I(x,t)W_I(t)$.
With eq.~(\ref{WI}) we have defined the free propagator $P_F(t)\equiv e^{-\ri
  K t}=\otimes_i e^{-\ri K_i t}\otimes e^{-\ri K_R t}\equiv \otimes_i
  P_i(t)\otimes P_R(t)$, and $L_I(x,t)=P_F(-t)L_I(x)P_F(t)$ is the
  interaction Liouvillian in the interaction picture.
We decompose furthermore $L_I=\sum_{k=1}^N
  L_{I,k}$, with $L_{I,k}X=[H_{I,k},X]$.  To second order in $L_I$
  we have    
\begin{eqnarray}
\langle A(t)\rangle&=&\tr\Big\{A\Big(P_F(t)-\ri\sum_k
\int_0^tP_F(t-t_1)L_{I,k}(x)P_F(t_1)\,dt_1\\
&-&\sum_{k,k'}\int_0^t dt_1\int_0^{t_1} dt_2
P_F(t-t_1)L_{I,k}(x)P_F(t_1-t_2) L_{I,k'}(x)P_F(t_2)+{\cal O}(L_{I,k}^3)
\Big)W(0)\Big\}\,.\nonumber
\end{eqnarray}
With an initial product state,
$W(0)=\otimes_{i=1}^N\rho_i(0)\otimes\rho_r(0)$, we obtain the zeroth
order term  $\langle A(t)\rangle_0\equiv \tr(AP_F(t)W(0))=\tr(\otimes_i
P_i(t)[\rho_i(0)]\otimes A P_R(t)[\rho_R(0)])=\tr_R(A\rho_R(t))$, as all
propagators are trace-preserving, and $\rho_R(t)\equiv 
P_R(t)[\rho_R(0)]$. Similarly, by 
explicitly writing $L_{I,k}X=\sum_\nu[S_{k,\nu}R_\nu,X]$, we obtain the
first order term
\begin{eqnarray} \label{A1d}
\langle A(t)\rangle_1&\equiv& -\ri\sum_k\tr\left(A
\int_0^tP_F(t-t_1)L_{I,k}P_F(t_1)\,dt_1W(0)\right)\,\\
&=&-\ri\sum_{k,\nu}\int_0^tdt_1\langle S_{k,\nu}(t_1)\rangle
\tr_R\left(AP_R(t-t_1)[R_\nu,\rho_R(t_1)]\right) 
\,,
\end{eqnarray}
where $\langle S_{k,\nu}(t)\rangle=\tr_k S_{k,\nu} P_k(t)[\rho_k(0)]$.
We
generalise the second simplifying assumption in Sec.~``Measuring the quantum
bus'' to
$P_R(t)[\rho_R(0)]=\rho_R(0)$, and $\tr_R(AP_R(t)[X])=f_A(t)\tr_R(AX)$ with
some function $f_A(t)$ \cite{Bonifacio71a}. This implies
that the initial state is decoherence-free
concerning the decoherence of $\clR$ alone. This is a natural
assumption for the state of an environment initially in thermal equilibrium,
or for a quantum bus in its ground state, such as an initially empty
cavity mode (see the example of superradiance).  We then have again $\langle
A(t)\rangle_1=0$.  
To second order in the interaction we find
\begin{eqnarray}
\langle A(t)\rangle&=&\langle A(0)\rangle_0\nonumber\\
&-&\sum_{\nu,\mu}\int_0^tdt_1\int_0^{t_1}dt_2\Big\{
N\Big( C_{S\mu\nu}^{(1)}(t_1,t_2)C_{AR\mu\nu}^{(1)}(t,t_1,t_2)-
C_{S\mu\nu}^{(2)}(t_1,t_2)C_{AR\mu\nu}^{(2)}(t,t_1,t_2)
\Big)\nonumber\\
&+&N^2 \langle S_{\mu}(t_1)\rangle \langle S_{\nu}(t_2)\rangle C_{AR\mu\nu}^{(3)}(t,t_1,t_2)
\Big\}\,,\label{Atd}\\
C_{AR\mu\nu}^{(1)}(t,t_1,t_2)&=&\tr_R\left(AP_R(t-t_1)[R_{\mu},P_R(t_1-t_2)[R_\nu\rho_R(t_2)]]\right)\\
C_{AR\mu\nu}^{(2)}(t,t_1,t_2)&=&\tr_R\left(AP_R(t-t_1)[R_{\mu},P_R(t_1-t_2)[\rho_R(t_2)R_\nu]]\right)\\
 C_{AR\mu\nu}^{(3)}(t,t_1,t_2)&=&C_{AR\mu\nu}^{(1)}(t,t_1,t_2)-C_{AR\mu\nu}^{(2)}(t,t_1,t_2)\\
C_{S\nu\mu}^{(1)}(t_1,t_2)&=&\tr_k\left(S_{\mu}P_k(t_1-t_2)[S_{\nu}\rho_k(t_2)]\right)-\langle
S_{\mu}(t_1)\rangle\langle S_\nu(t_2)\rangle\\  
C_{S\nu\mu}^{(2)}(t_1,t_2)&=&\tr_k\left(S_{\mu}P_k(t_1-t_2)[\rho_k(t_2)S_{\nu}]\right)-\langle
S_{\mu}(t_1)\rangle\langle S_\nu(t_2)\rangle\,,
\end{eqnarray}
where $\rho_R(t)=P_R(t)[\rho_R(0)]$. The index $k$
is arbitrary, $k=1,\ldots,N$, as we have assumed all systems $\cS_k$
identical and identically prepared.  All $x$ dependence is in the operators
$S_\nu$. Equation~(\ref{Atd}) also gives $\langle 
A^2(t)\rangle$ by replacing $A\to A^2$, and $\langle
A(t)\rangle^2$ to order ${\cal O}(H_I^2)$. The equation obtained by
inserting these expressions into eq.~(\ref{dxdef}) generalizes the result
(\ref{dxFin}) to decoherence on top of the unitary evolution considered in
Sec.~``Measuring the quantum bus''. We see that the basic structure of the
result for $\delta 
x$, and in particular its scaling with $N$ is
unchanged, but the expectation values and correlation functions are replaced
by more complicated expressions involving in general mixed states and
non-unitary evolution of individual subsystems. 

\subsection*{Quantum parameter estimation for a Markovian master equation}
In standard descriptions of decoherence one traces out the heat bath and gets
a master equation for the reduced density matrix $\rho_s$ of $\cS$
alone. Using quantum parameter estimation
theory generalized to non-unitary evolution we now show that for 
Markovian decoherence with an initially
decoherence-free state, measuring an arbitrary 
$x$-independent observable on $\cS$ alone gives at best a $\delta x_{\rm
  min}\sim 1/\sqrt{N}$. This corroborates
the result found for unitary evolution that the important quantum system to
measure is the common quantum bus $\clR$, rather than $\cS$. 

The Markovian master equation for $\rho_s(t)$
obtained by tracing out $\clR$  has the Lindblad-Kossakowski form   
\begin{eqnarray} \label{Lgen}
\dot{\rho_s}(t)&=&{\Lambda}(x)[\rho_s(t)]\equiv
\gamma\sum_{\alpha=1}^d\left([F_\alpha(x),\rho_s(t) 
  F_\alpha^\dagger(x)]+h.c.\right)\,,
\end{eqnarray}
where we work in the interaction picture and assume that there is no
additional unitary evolution.
The $F_\alpha(x)$ are arbitrary linear (not
necessarily hermitian) operators which have inherited the $x$-dependence from
the interaction  hamiltonian $H_I(x)$, and $d$ is the total number of
generators. 
Note that we can restrict ourselves to an
  initially pure state, as for any linear propagation one cannot do better
  with a mixed state than with the pure states from which it is mixed
  \cite{Braun10}. We expand the
  Markovian time evolution to first order in $t$, $\rho=
  |\psi\rangle\langle \psi|+t\, {
  \Lambda(x)}|\psi\rangle\langle \psi| +{\cal O}(t^2)$, and linearise 
  $F_\alpha(x)$ about the value of $x$ where we want to measure. We
  set that value, without restriction of generality, to zero,
  i.e.~$F_\alpha(x)=x F_\alpha'+F_\alpha(0)$, and assume that the initial
  state is decoherence-free at $x=0$. The Bures distance can still
  be evaluated in a straight-forward fashion as the state at $x=0$ remains
  pure.
  One finds
$ds^2\left.\right|_{x=0}=8\gamma t\sum_\alpha
K_{|\psi\rangle}(F_\alpha^\dagger,F_\alpha)\,dx^2$.
As a consequence, the ultimate quantum limit of the sensitivity with
which the parameter $x$ can be estimated from the parametric
dependence of the master equation, starting from a pure state
$|\psi\rangle$, reads 
\begin{equation} \label{dxfin}
\delta x_{\rm min}=
\frac{1}{2\sqrt{2M\gamma
  t}
  \left(\sum_{\alpha=1}^dK_{|\psi\rangle}(F_\alpha^\dagger,F_\alpha)\right)^{1/2}}\,.  
\end{equation}
With $F_\alpha=\sum_{r=1}^NF_{\alpha,r}$ we
obtain
$
K_{|\psi\rangle}(F_\alpha^\dagger,F_\alpha)=\sum_{r,s=1}^NK_{|\psi\rangle}(F_{\alpha,r}^\dagger,F_{\alpha,s})$.
For an initially entangled state,
$K_{|\psi\rangle}(F_\alpha^\dagger,F_\alpha)$ can be of order $N^2$.
This can be seen from the example of the GHZ state
$|\psi\rangle=(|0\ldots 0\rangle+|1\ldots 1\rangle)/\sqrt{2}$, and a single
generator $F_{1, 
  i}=\sigma^{(i)}_{z}$ with $\sigma^{(i)}_{z}$ the Pauli
$z$ matrix for subsystem $i$. Then 
$K_{|\psi\rangle}(F_1^\dagger,F_1)=N^2$, and one obtains a $1/N$ scaling of
$\delta x_{\rm min}$, just as in the case of unitary evolution.
However, if the initial state factorizes,
$|\psi\rangle=\otimes_{l=1}^N|\varphi_l\rangle$, there are no
correlations between different subsystems $r$ and $s$, and we thus
have only the sum of correlations in all subsystems,
$K_{|\psi\rangle}(F_\alpha^\dagger,F_\alpha)=\sum_{s=1}^NK_{|\varphi_s\rangle}(F_{\alpha
  ,s}^\dagger, F_{\alpha,s})$, which is at most of order $N$, and $\delta
x_{\rm min}$ scales as $1/\sqrt{N}$, again just as in the case of unitary
evolution. This shows once more that a measurement of an $x$-independent
observable on $\cS$ does not allow to do better than in the standard
situation of unitary evolution
of $\cS$ without coupling to a common quantum bus. 

Interestingly,
superradiance {\em is} 
described by a master equation of $\cS$ alone after tracing out the cavity
mode.  But a measurement on $\clR$ (the
number of photons in the cavity) translates in that case to a measurement on
the $\cS_i$ that depends itself on $x$.  In this way it is still possible to
achieve a $1/N$ scaling of $\delta x$.\\

The authors declare to have no competing financial interests.\\

Author contributions: DB conceived the original idea, worked out the
general formalisms and examples,  found the underlying quantum-information
theoretical principle, and  
wrote the manuscript. JM and DB both
calculated the specific example of superradiance and discussed all
aspects of the work at all stages.\\

\bibliography{../../mybibs_bt}

{\em Acknowledgments:} We thank Peter Braun, David Guery-Odelin,
Jacob Taylor, and Eite Tiesinga for useful
discussions. This work was supported by the Agence
National de la Recherche (ANR), project INFOSYSQQ. Numerical calculations
were partly performed at CALMIP, Toulouse. J.M.\ thanks the Belgian
F.R.S.-FNRS for financial support, and DB thanks the Joint Quantum
Institute, where part of this work was done, for hospitality.

Correspondence and requests for additional material should be
addressed to DB.

\pagebreak

\pagebreak

\part*{\center \large Supplementary Discussion for the manuscript ``Heisenberg-limited sensitivity with decoherence-enhanced  measurements''}
\vskip 15pt

\makeatletter %
\def\thefigure{{Supplementary Figure S}\@arabic\c@figure}
\def\fnum@figure{\textbf{\thefigure}}
\makeatother %

\setcounter{figure}{0}
\setcounter{equation}{0}
\renewcommand{\theequation}{S\arabic{equation}}

\sloppy

\section*{\large Pure interaction and decoherence}
We reconsider here the simple interaction-dominated model (26), but
with additional decoherence of 
the subsystems.  To
simplify things and make connection to the language of quantum information
theory, we consider directly the case where both the $\cS_i$ and $\clR$ are
qubits. In the eigenbasis of $S_i$ and $R$, the interaction $H_I(x)$ can
be written as 
$H_I(x)=x\sum_{i=1}^N\sigma_{z}^{(i)}\otimes\sigma_{z}^{(0)}$, where we use
the 
label 0 for $\clR$. As observable on $\clR$ we chose $A=\sigma_x$.  Without
decoherence one obtains the uncertainty $\delta 
x=1/(2\sqrt{M}Nt)$ in 
agreement with (28). We will restrict ourselves to phase-flip
channels acting independently on all qubits with a rate $\Gamma$,
\begin{eqnarray}
\Lambda_{\sigma_z^{(i)}}X&\equiv&\frac{\Gamma}{2}\left([\sigma_z^{(i)}X,\sigma_z^{(i)}]+[\sigma_z^{(i)},X\sigma_z^{(i)}]\right)\,.\end{eqnarray}
As before we decompose the Liouvillian
$L_I(x)X=[H_I(x),X]=\sum_{i=1}^N L_{I,i}X$. One easily verifies
$[\Lambda_{\sigma_z^{(i)}},L_{I,k}]=0$ for all $i,k$.

Consider first phase flips of the qubits $\cS_i$.
The equation of motion for the full density matrix $W(t)$,
$\dot{W}(t)=\sum_{i=1}^N(L_{I,i}+\Lambda_{\sigma_z^{(i)}})W$,  is
easily 
solved in the computational basis, ($\ba=a_Na_{N-1}\ldots a_0$, $a_i=\pm 1$,
$i=0,\ldots,N$), 
$W_{\ba\bb}(t)=W_{\ba\bb}(0)\exp\left(-\ri\sum_{i=1}^N\left(x\left(a_ia_0-b_ib_0\right)+\ri\Gamma(a_ib_i-1)\right)t\right)$.
The reduced density matrix for $\clR$ alone 
is therefore 
\begin{equation} \label{rhoab}
\rho_{a_0b_0}=\sum_{a_1,\ldots,a_N=\pm1}W_{a_N\ldots a_1a_0a_N\ldots
  a_1b_0}(0)\exp\left(-\ri\sum_{i=1}^Nxa_i(a_0-b_0)t\right)\,.  
\end{equation}
We see that $\Gamma$ drops out completely, and dephasing of the $\cS_i$ has
therefore no impact at all on the total accumulated phase of $\clR$, and
therefore on $\delta x$.

If phase flips occur in $\clR$,
$\dot{W}=\left(\left(\sum_{i=1}^NL_{I,i}\right)
+\Lambda_{\sigma_z^{(0)}}\right)W$,   
  we have
  $W_{\ba\bb}(t)=W_{\ba\bb}(0)\exp\left(-\ri\left(\sum_{i=1}^Nx\left(a_ia_0-b_ib_0\right)+\ri  
\Gamma(a_0b_0-1)\right)t\right)$. For the initial state
$|\psi_0\rangle=\ket{+\ldots +}(\ket{+}+\ket{-})/\sqrt{2}$, 
we find the reduced density matrix element 
$\rho_{+-}(t)=\exp(-2\Gamma t)\cos(2N x t)$. Thus, the decay of
$\langle \sigma_x(t)\rangle$ is independent of $N$, and only the
prefactor in the scaling of $\delta x$ with $N$ changes.

\section*{\large Superradiance: practical issues}\label{sec.SR}
We now address several additional technical questions that are important for
a possible 
experimental implementation of our method of measuring the length of a
cavity using decoherence-enhanced measurements. 
\subsection*{Preparation of initial state}
 In order to prepare the product state
$|\psi_0\rangle=\left(\otimes_{i=1}^N\frac{1}{\sqrt{2}}(|t_-\rangle_i+|s\rangle_i)\right)|0\rangle$
 it is 
 helpful to use three--level atoms with a lambda 
structure.  Let $|0\rangle$ and the additional state $|2\rangle$ be
hyperfine (HF) states, and assume that their energies are sufficiently split
such that only the transition 
$|0\rangle\leftrightarrow
|1\rangle$ resonates with the cavity mode. We assume further that the second
optical lattice can be moved along the cavity axis, such that controlled
pairwise collisions of corresponding
atoms in the two lattices can be induced.  Entangled pairs of atoms in their
HF split ground states can thus
be created (for atoms in the same lattice
this has been demonstrated experimentally, see Supplementary Reference [43]
 for a 
review). After the creation of an entangled HF state $|\psi'_0\rangle$,
that differs from  $|\psi_0\rangle$ by the replacement of
states 
$|1\rangle$ by states $|2\rangle$, the second lattice is moved back to its
original position.  Now one can
selectively excite the $|2\rangle$ states by a
laser pulse in resonance with the $|2\rangle\leftrightarrow
|1\rangle$ transition, that replaces the singlets in the (very long
lived) HF states by the desired
singlets of
the $|0\rangle$ and $|1\rangle$ states and thus produce
$|\psi_0\rangle$. However, as such, the method is not of much
practical use yet, as it will be virtually impossible to park the second
lattice at the exact position corresponding to coupling constants which
render $|\psi_0\rangle$ decoherence free. The extreme
sensitivity of the collective decoherence with respect to changes of the
coupling constants plays against us here, and will lead to leaking of light
from the cavity after the excitation 
$|\psi'_0\rangle\to |\psi_0\rangle$, if the exact position
corresponding to $|\psi_0\rangle\in$ DFS is not achieved. But it is
possible to
position the second lattice at the required position with a precision of
${\cal O}(1/N)$
using a feed-back
mechanism and a part of the quantum resources. With the atoms in the state
$|\psi'_0\rangle$, do the following repeatedly in order to find the
optimal position: Excite a part of the entangled HF pairs containing ${\cal
  O}(N/\ln
N)$ atoms with the laser, measure $\langle n_{\rm ph}\rangle$, and use the
measurement results to bracket
the minimum of $\langle n_{\rm ph}\rangle$ as function of the
lattice position. The minimum of $\langle n_{\rm ph}\rangle$ indicates that
the position
corresponding to the DFS is achieved. Using golden section search, the
minimum  can be bracketed to precision $1/N$
in ${\cal O}(\ln N)$
moves, as at each step the sensitivity of the measurement of the
position of the
lattice is of order ${\cal O}(\ln N/N)\sim {\cal O}(1/N)$. Once the minimum
is found, excite the
remaining unused pairs (there
should be still a number of pairs of ${\cal O}(N)$) to the desired state
$|\psi_0\rangle$.  That state is now decoherence-free, and the system ready
to detect small changes of the position of one of the mirrors. Note that for
this method it is not necessary to know which exact state is produced
in the controlled collisions and subsequent laser excitation.

\subsection*{Imperfections} 
Here we discuss how the most important
additional noise sources affect the photon statistics.  In view of
eqs.~(20,21), we 
restrict ourselves to analyzing
$\alpha=-\langle\dot{J}_z(x,0)\rangle\equiv\tr (J_z
L_I(x)[\rho_s(0)])=2\gamma\langle 
J_+J_-(x,0)\rangle$, where $J_z\equiv  
\frac{1}{2}\sum_{i=1}^N\sigma_z^{(i)}$ is the total population inversion of
the atoms. Energy conservation shows that $\alpha$ can be interpreted as
the initial photon escape rate from the cavity.

\subsubsection*{Spontaneous emission}\label{sec.sponem}
It is easily verified that the spontaneous emission term 
in eq.~(17) 
leads to a contribution to $\langle
\dot{J}_z(0)\rangle$ that scales as
${\cal O}(N)$, to be compared to the term of ${\cal O}(N^2)$ from collective
emissions, 
eq.~(21). 
 The modification of the
initial 
state due to spontaneous emission is not an issue, as we consider an
initial product state.  Atoms which decay become simply unavailable
for collective decoherence, but since their number is proportional to
$N$ this does not change the scaling of $\delta x$ with $N$. Also, note that 
spontaneous emission sends photons into the entire open space but {\em not}
into the cavity, whereas the collective emission escapes {\em exclusively}
through the leaky cavity mirror.  Therefore, in addition, the two
contributions 
can be well separated experimentally by observing only the photons
which escape through the cavity mirror. 

\subsubsection*{Fluctuating coupling constants}
Another obvious concern are fluctuations of the coupling constants.
In order to reduce the noise of the measurement of  $\langle n_{\rm
  ph}\rangle$, one may want to repeat the experiment $M$ times with $M\gg
1$. The exact coupling constants might
fluctuate during the
averaging, e.g.~due to fluctuating traps caused by vibrations in the
set up. But even for perfectly stable traps, thermal motion,
or even quantum fluctuations in the traps will lead to fluctuating
$g_i$. We now show that the cost in sensitivity
of these fluctuations depends on 
their {\em correlations}. Fluctuations correlated between pairs of
atoms $l$ and $l+N/2$ come at no cost, completely uncorrelated
fluctuations lead back to a $\delta x\propto 1/\sqrt{N}$, and fluctuations
perfectly correlated within 
the same lattice but not between the two lattices lead to noise
indistinguishable from 
the signal. 

To see this, consider fluctuations
$\delta{g}_i$ of the
${g}_i$ about their mean values ${G}_I$,
${g}_i={G}_1+\delta{g}_i$ for $i=1,\ldots,N/2$,
${g}_i={G}_2+\delta{g}_i$ for $
i=N/2+1,\ldots,N$. The generator $J_-$ reads then
$J_-=\frac{1}{g}\sum_{i=1}^N{g}_i\sigma_-^{(i)}$. We 
introduce the correlation matrix $C_{ij}=\overline{\delta
  {g}_i\delta {g}_j}/g^2$, where the over-line denotes an average
over the ensemble describing the fluctuations, and assume $\overline{\delta
  g_i}=0$ for simplicity.
Then
$J_-|\psi_0\rangle=-\frac{1}{2g}\sum_{i=1}^{N/2}({g}_i-{g}_{i+\frac{N}{2}})|t_-\rangle_i\bigotimes_{l\ne
  i}^{N/2}|\varphi_l\rangle_l$ and
\begin{eqnarray}
\alpha&=&
\frac{\gamma}{2g^2}\Big(\sum_{i=1}^{N/2}
  ({g}_i-{g}_{i+N/2})^2+\sum_{\stackrel{i,j=1}{i\ne
      j}}^{N/2}\frac{1}{2}\left({g}_i-{g}_{i+N/2}\right)\left({g}_j-{g}_{j+N/2}\right) 
\Big)\,.\label{outdfs}
\end{eqnarray}
Equation~(\ref{outdfs}) predicts a background $\alpha_{\rm bg}$ in
the photon escape rate on top of the value for $\delta g_i=0$,
\begin{equation} \label{a0}
\alpha_0=\frac{\gamma}{4}x^2(2N+N^2)\,,
\end{equation}
with $x=(G_1-G_2)/(2g)$, see 
eq.~(21), and 
fluctuations 
$\delta \alpha_f$, $\alpha=\alpha_0+\alpha_{\rm bg}+\delta
\alpha_f$, where the average background is given by
\begin{eqnarray}
\overline{\alpha}_{\rm
  bg}&=&\frac{\gamma}{2}\Big\{\sum_{i=1}^{N/2}{\cal C}_{ii}
  +\frac{1}{2}\sum_{i\ne j}^{N/2}{\cal C}_{ij}\Big\}\nonumber\,,
\end{eqnarray}
where ${\cal C}_{ij}\equiv \Big( 
C_{ij}+
C_{i+\frac{N}{2}\,j+\frac{N}{2}}
-C_{i\,j+\frac{N}{2}}
-C_{i+\frac{N}{2}\,j}
\Big)$.
The average background $\overline{\alpha}_{\rm bg}$ can be determined
independently at $x=0$, and subtracted from the
signal; it does therefore not influence the sensitivity of the
measurement. The remaining noise $\delta\alpha_f$ fluctuates about zero,
\begin{eqnarray}
\delta\alpha_f&=&x\gamma\left(\frac{N}{2}+1\right)\sum_{i=1}^{N/2}\frac{\Delta
g_i}{g}\,,\label{flu}
\end{eqnarray}
where $\Delta{g}_i\equiv \delta{g}_i-\delta{g}_{i+N/2}$.
Its standard deviation
$(\overline{\delta
\alpha_f^2})^{1/2}=\gamma|x|D$,
where $D=(\frac{N}{2}+1)\left(\sum_{i,j=1}^{N/2}{\cal
  C}_{ij}\right)^{1/2}$  
 translates into additional noise in the
number of detected photons.
Several interesting cases can be considered:
\begin{enumerate}
  \item Fully uncorrelated fluctuations, $C_{ij}=C_i\delta_{ij}$, where
  $\delta_{ij}$ stands for the Kronecker-delta: Here we get
  $D=(\frac{N}{2}+1)\left(\sum_{i=1}^{N/2}(C_i+C_{i+N/2})\right)^{1/2}$, which
  is in 
  general of order $N^{3/2}$, and leads back to the standard quantum limit,
  $\delta x\sim 1/\sqrt{N}$ for large $N$.

\item Pairwise identical fluctuations between the two sets:
  $C_{ij}=C_{i+\frac{N}{2}\,j}=C_{i
  \,j+\frac{N}{2}}=C_{i+\frac{N}{2}\,j+\frac{N}{2}}$ for $
  i,j=1,\ldots,N/2$. This can be the consequence of fully correlated
  fluctuations, $C_{ij}=C \,\,\,\forall\,\,\, i,j$.  Alternatively, such a
  situation arises for
  example for atoms initially
  arranged symmetrically with respect to an anti-node such
  that ${G}_1^{(0)}={G}_2^{(0)}$, if
  the two atoms
  (or ions) in each pair $l$ ($l=1,\ldots,N/2$) are locked into a common
  oscillation. This should be the case for two trapped ions repelling each
  other through a strong Coulomb interaction, and cooled below the
  temperature corresponding to the frequency
  of the breathing mode. Equation~(\ref{flu}) then
  gives $\delta\alpha_f=0$, i.e.~no additional noise from the fluctuations
  of the couplings. Note, however, that for initial
  ${G}_1^{(0)}\ne    {G}_2^{(0)}$
  the more general DFS leads to a more complicated condition for the
  correlations,
  $C_{ij}|{G}_2^{(0)}|^2
+C_{i+\frac{N}{2}\,j+\frac{N}{2}}|{G}_1^{(0)}|^2
-C_{i\,j+\frac{N}{2}}{G}_2^{(0)}{G}_1^{(0)}
-C_{i+\frac{N}{2}\,j}{G}_1^{(0)}{G}_2^{(0)}=0$, which might be
  harder to achieve.
\item Correlated fluctuations within a set, but uncorrelated between the two
  sets, $C_{ij}=C$ for $i,j\in \{1,\ldots,N/2\}$ or $i,j\in \{N/2+1,\ldots,
  N\}$, but $C_{ij}=0$ for 
  $i\in \{1,\ldots,N/2\}$ and $j\in \{N/2+1,\ldots,
  N\}$ or vice versa. This case leads to a noise of order ${\cal O}(N^2)$,
  the 
  worst case scenario. However, this comes as no surprise, as such
  correlations are indistinguishable from the signal: all the atoms in a
  given set move in a correlated fashion, but independently from the atoms
  of the other 
  set. This modifies the couplings in the same fashion as if the length of
  the cavity was changed. 
\end{enumerate}

Case (2) above is clearly the most favorable situation. If there are
no other background signals depending on $N$, we keep the $1/N$
scaling of $\delta x$. In order to favor case (2) over cases
(1),(3), it appears to be advantageous to work with ions and to try
to bring the ions in a pair as closely together as possible, thus
strongly correlating their fluctuations, while separating the ions
in the same set as far as possible.

\subsubsection*{Imperfections in initial state preparation} 
Suppose that instead of the state $|\psi_0\rangle$, the state of the atoms
\begin{equation} \label{psi1}
|\tilde{\psi}_0\rangle=\bigotimes_{l=1}^{N/2}|\varphi\rangle_l\mbox{
with }
|\varphi\rangle_l=a|t_-\rangle_l+b|s\rangle_l+c|t_0\rangle_l+d|t_+\rangle_l
\end{equation}
was prepared (we consider the same state for all pairs for
simplicity, but this is not essential, and assume the state normalized). Then,
\begin{eqnarray}
\alpha&=&\gamma\Big[ \frac{N}{2}
\Big(\big((c-b){G}_1+(c+b){G}_2\big)^2+2d^2({G}_1^2+{G}_2^2)\Big)\nonumber\\
&&+\frac{N}{2}\left(\frac{N}{2}-1\right)
\Big(({G}_2-{G}_1)b(a-d)+({G}_2+{G}_1)c(a+d)\Big)^2
\Big]/g^2\,.
\end{eqnarray}
The derivative of $\alpha$ with respect to ${G}_2$
is of order
${\cal O}(N^2)$, and thus still allows to find the minimum of $\alpha$ as
function of
the position of the second lattice with a precision of order ${\cal
  O}(1/N)$. At the minimum a
component outside the DFS persists, such that photons will leak out of the
cavity, but the average rate is only of order ${\cal
  O}(N)$, and can be measured separately and subtracted from the
signal. Changes $\delta{G}_2$ of  ${G}_2$ away from the
position of the minimum still lead to a signal that scales, for
large $N$ as $N^2$,
$\alpha=\frac{\gamma}{4}\left(a(c+b)+d(c-b)\right)^2N^2\delta{G}_2^2/g^2$, and
the analysis leading to the $1/N$ scaling of $\delta x$ still
applies. 

\begin{figure}[h!t]
\epsfig{file=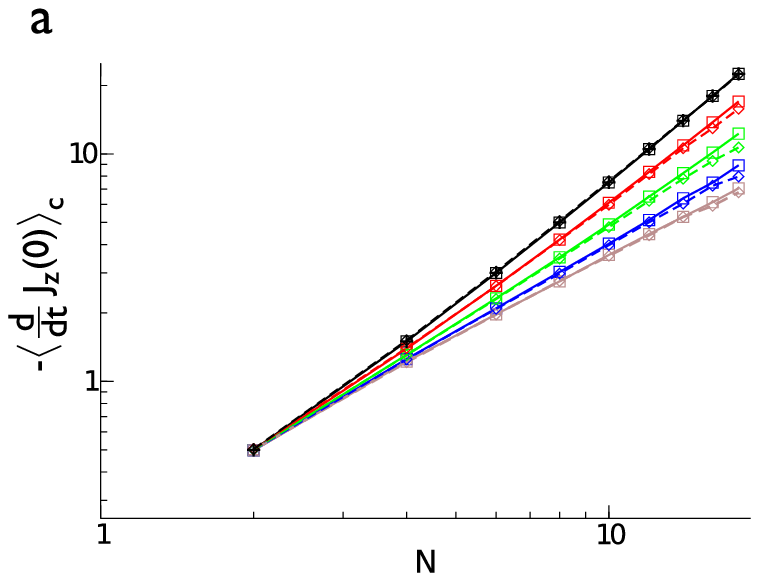,width=0.4\textwidth,angle=0}\hspace{1.5cm}
\epsfig{file=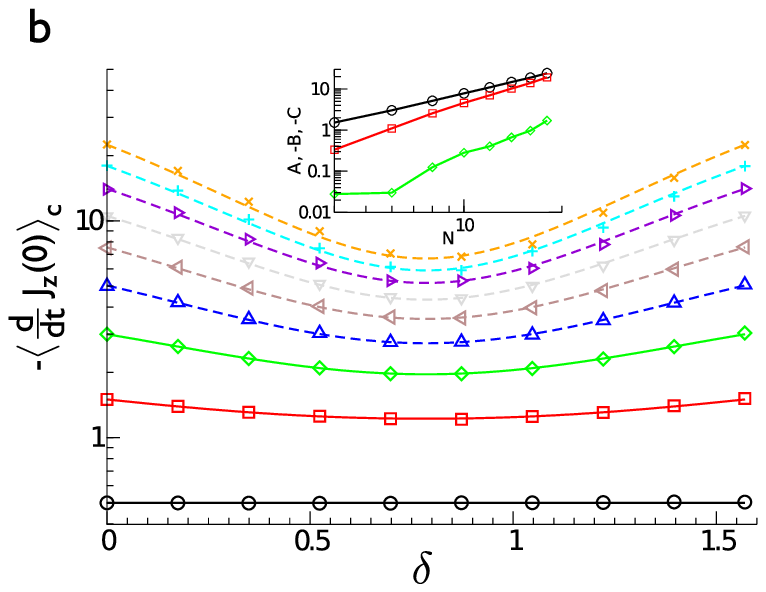,width=0.4\textwidth,angle=0}\hspace{0.3cm}
\caption{Effects of imperfect state preparation. {\bf (a)} Scaling of collective
  photon escape rate $\alpha=-\langle
\dot{J}_z(x,0)\rangle$ (in
units
  $\gamma|{G}_1-{G}_2|^2/g^2$) in
the mixture of DFS states reached by relaxation from state
$|\tilde{\psi}_0\rangle$, eq.~(\ref{psi1}). Data for
$\delta=\delta_0$ ($\delta=\pi/2-\delta_0$) denoted by squares and
full lines (diamonds and dashed lines), respectively;
$\delta_0=0,\pi/9,2\pi/9,3\pi/9$ and $4\pi/9$ in black, red, green,
blue, and brown.  Exact analytical results for $\delta=0$,
eq.~(21) 
shown with black crosses.  Full and dashed lines
are guides to the eye only. {\bf (b)} Dependence of $\alpha$ on imperfection
parameter $\delta$.
$N=2$, 4,6,8,10,12,14,16,18 in black circles, red squares, green
diamonds, blue triangles up, brown triangles left, grey triangles
down, violet triangles right, cyan pluses, orange Xs, respectively.
The full lines for $N$=2,4,6 are exact analytical results. The
dashed lines for $N=8,\ldots,18$ are fits to
$A+B\sin(2\delta)+C\cos(4\delta)$. The inset shows the scaling of
the coefficients $A$ (black circles), $-B$ (red squares), and $-C$
(green diamonds) as function of $N$.}\label{fig.zdot}
\end{figure}

Alternatively, one can get rid of the additional background by letting
the system relax before measuring changes of $L$. Indeed, any state
with a component outside the DFS will relax to a DFS state or mixtures of
DFS states within a time of order 
$1/\gamma$ or less. Components with large total pseudo-angular
momentum $J$ relax in fact in much shorter time of order $1/(J\gamma)$.
The DFS states reached through relaxation starting from
$|\tilde{\psi}_0\rangle$ still allow a scaling of $\alpha$ close to 
$N^2$. We have shown this by simulating the relaxation process with
the help of the stochastic Schr\"odinger equation (24).
Using an
Euler scheme with a time step of 
$0.01/\gamma$, we followed the convergence of $\psi(t)$ to DFS
states for states with
$(a,b,c,d)=(\cos\delta,1,0,\sin\delta)/\sqrt{2}$, until the norm of
the difference $|\psi(t+dt)\rangle-|\psi(t)\rangle$ dropped below
$10^{-12}$. In these final dark states, randomly distributed over
the DFS, we calculated $\alpha$, and averaged over a large number
$n_r$ of realizations of the stochastic process ($n_r=10^5$, $10^4$,
$10^4$, $2.5\cdot 10^3$, $10^3$, $1.25\cdot 10^3$, 250, 200, and 250
for $N=2,4,6,8,10,12,14,16$, and $18$). \ref{fig.zdot} shows
the scaling of $\alpha$ as function of $N$ for different values of
$\delta$ for $0\le\delta\le\pi/2$ up to $N=18$. Within this
numerically accessible range of $N$, $\alpha$ follows a power law
$\alpha\propto N^p$ with an exponent $p$ that decays only gradually
with $\delta$ for $\delta\le \pi/4$. Moreover, that decay might be a
finite size effect: Note that, surprisingly, $\alpha(\delta)$
appears to be close to symmetric with respect to $\delta=\pi/4$.
This is corroborated by exact analytical calculations based on the
diagonalisation of $L_I(x)$, which lead to $\alpha=1/2$ for
$N=2$, $\alpha=(55-12\sin(2\delta)-\cos(4\delta))/36$ for $N=4$, and
$\alpha=(303-110\sin(2\delta)-3\cos(4\delta))/100$ for $N=6$ (in
units $\gamma|{G}_1-{G}_2|^2/g^2$). The plot shows that all
numerical data can be very well fitted by
$\alpha=A+B\sin(2\delta)+C\cos(4\delta)$. From 
eq.~(21) 
we know that $A+C$ has to scale as $N^2$ for sufficiently large $N$.
Both $B$ and $C$ are negative for all $N$ for which we have data,
and $C$ appears to be negligible. \ref{fig.zdot} shows that
$-B$ increases even more rapidly than $N^2$ (a fit in the range
$N=8,\ldots,18$ gives a power law $N^{2.4}$). But $B$ has to cross
over to a power law $N^p$ with $p\le 2$, unless other Fourier
components start contributing significantly. Otherwise, $\alpha$
would become negative for $\delta>0$. This indicates that for large
$N$ the scaling of $\alpha$ is in fact $N^2$ for all $\delta$. 

In summary, our method still works, even if the product state
$|\psi_0\rangle$ is not prepared perfectly. One has the choice to start
measurement immediately after state preparation, which gives an
additional background of order $N$, or to wait a time of the order
of a few $1/\gamma$ after preparation of the initial state, until no
more photons leave the cavity through the mirror, with no additional
background penalty. In both cases the scaling of the uncertainty of $\delta
L/L$ in  a subsequent measurement of a
small change of $L$ is still $\propto 1/N$ .

\section*{\large Supplementary References}
\noindent [43]~~\parbox[t]{\textwidth}{Bloch, I., Quantum coherence and entanglement with ultracold atoms in optical\\ lattices, \emph{Nature} \textbf{453}, 1016--1022 (2008).}

\end{document}